# High frequency magneto-acoustic resonance through strain-spin coupling in perpendicular magnetic multilayers

De-Lin Zhang[1†], Jie Zhu[2†§], Tao Qu[3†], Dustin M. Lattery[2†], R. H. Victora[1,3]*, Xiaojia Wang[1,2]* and Jian-Ping Wang[1]*

[1]Department of Electrical and Computer Engineering, University of Minnesota, MN 55455, USA;
[2]Department of Mechanical Engineering, University of Minnesota, MN 55455, USA;
[3]School of Physics and Astronomy, University of Minnesota, MN 55455, USA;
[†]These authors have equal contributions to this work.
[§]Present address: Key Laboratory of Ocean Energy Utilization and Energy Conservation of Ministry of Education, Dalian University of Technology, Dalian, Liaoning 116024, China.
*Author to whom correspondence should be addressed: jpwang@umn.edu, wang4940@umn.edu and victora@umn.edu

**It is desirable to demonstrate experimentally an extremely high resonant frequency, assisted by strain-spin coupling, in technologically important perpendicular magnetic materials for device applications. Here we directly observe the coupling of magnons and phonons in both time and frequency domains upon femtosecond laser excitation. This strain-spin coupling leads to a magneto-acoustic resonance in perpendicular magnetic [Co/Pd]$_n$ multilayers, reaching frequencies in the Extremely-High-Frequency (EHF) band, *e.g.*, 60 GHz. We propose a theoretical model to explain the physical mechanism underlying the strain-spin interaction. Our model explains the amplitude increase of the magneto-acoustic resonance state with time and quantitatively predicts the composition of the combined strain-spin state near the resonance. We also detail its precise dependence on the magnetostriction. The results of this work offer a potential pathway to manipulating both the magnitude and timing of extremely high-frequency and strongly coupled magnon-phonon excitations.**

**Teaser sentence:** Time-resolved magneto-acoustic resonance demonstrates strong magnon-phonon coupling at extremely high frequency.

# INTRODUCTION

The exploration of innovative approaches and new physics to couple energy transfer between magnons and phonons has recently renewed interest in both the scientific and technological communities (*1-2*). Among the abundant physical phenomena (*3-7*) arising from the magnon-phonon coupling, the strain-assisted magneto-acoustic resonance in ferromagnetic (FM) materials provides an energetically efficient path for rapid switching of spin state, which is required for applications in cloud storage, advanced spin memory, logic and other spintronic devices (*8-10*).

Strain can be launched by an ultrafast optical excitation, specifically femtosecond laser pulses, through thermal expansion (*2,11-15*). The brief increase in thermal energy results in a local strain. This strain travels through the FM layers in the form of acoustic waves, with frequencies tunable by both the material composition and the layer geometry. In addition, the femtosecond laser pulses excite thermal demagnetization and thus introduce magnons (*16*). The acoustic waves act directly within the FM layer *via* magnetostriction (or the Villari effect) to couple with the magnetization. Thus, by simultaneously generating strain and spin within the ferromagnetic layer, the ultrafast optical approach provides a powerful tool to initiate the coupling between phonons and magnons. The femtosecond time scale of the laser pulse is smaller than the relaxation time scales of phonons and magnons, which allows the capture of the magnon-phonon coupling processes in the time domain, such as strain-assisted large-angle magnetization precession (*2*). In contrast, efforts have also been devoted to launching strain electrically by using piezoelectric materials (*17-19*), which induce a Surface Acoustic Wave (SAW) in the substrate beneath the FM layer. In such experiments, the SAW propagates to the FM layer and modifies its FM properties by changing the lattice spacing, leading to a different phonon propagation direction.

Most experimental and theoretical studies of the magnon-phonon coupling in literature are focused on materials unsuitable for spintronic applications due to their low resonant frequency or low Curie temperature, such as dilute magnetic semiconductor materials (*20-24*) and metallic FM materials (*e.g.*, Terfenol-D (*2*),

Gafenol (*25,26*), Bi-YIG (*27*) and Ni (*28-31*)). To date, direct experimental demonstration remains elusive for the magnon-phonon coupling, as well as ultrahigh frequency magneto-acoustic resonance in materials with perpendicular magnetic anisotropy (PMA). To remedy this, we choose [Co/Pd]$_n$ multilayers as a model system in this work. These materials possess high PMA at room temperature and a relatively large magnetostriction coefficient $\lambda$ ($\sim -1 \times 10^{-4}$) (*32,33*), holding great potential for various technological applications (*34-38*). Thus, they serve as an ideal platform for investigating the magnon-phonon coupling.

Here we report an extremely high frequency (EHF) magneto-acoustic resonance up to 60 GHz in the PMA [Co/Pd]$_n$ multilayer, originating from strong magnon-phonon coupling following excitation by femtosecond laser pulses. The resonance shows an enhanced wave envelope in the time domain, an anti-crossing in frequency domain, and significant mixing of both magnons and phonons as predicted by a coupled model. Such an observation is also demonstrated by both ultrafast measurement signals and micromagnetic simulation. These observations all indicate that a hybridized quasiparticle comprised of both magnon and phonon exists in perpendicular magnetic multilayers, whereby the energy transfers among magnon and phonon systems efficiently. This allows the acoustic wave to substantially influence the magnetization at an ultrafast picosecond scale. Thus, our work paves a potential pathway to enabling an EHF magneto-acoustic resonance through the magnon-phonon coupling and suggests a possibility of ultrahigh-speed strain-assisted magnetization switching in a technologically relevant magnetic system.

## RESULTS

**Sample characteristics and ultrafast instrumentation.** Figure 1A illustrates the time-resolved magneto-optical Kerr effect (TR-MOKE) and time-domain thermoreflectance (TDTR) ultrafast measurement configuration, together with a schematic of the sample stack. The thermoreflectance signals collected in TDTR measurements are proportional to the intensity change in the reflectivity of the probe beam, which contain information of both temperature and acoustics (*39,40*). Thus, the TDTR signals represent strain waves in the sample. The TR-MOKE method employed here is a pump-probe technique capable of recording the time evolution of magnetization (**M**), through the change in the polarization of a probe beam reflected from FM films (*39*). Typically, raw signals from TR-MOKE measurements may contain non-MOKE components, whose contributions are found to be negligible in determining the frequencies and relaxation times of spin precession and acoustic waves. More details about the TDTR and TR-MOKE experiment techniques are provided in Supplementary Note 1.

For a typical polar TR-MOKE measurement, a damped oscillation feature is expected in the signals due to the spin precession initiated by the rapid temperature rise from pump excitation, when an external magnetic field ($H_{ext}$) is applied (*39*). However, the TR-MOKE signals usually do not show magnetization oscillations in the absence of $H_{ext}$, since the equilibrium axis of the spin precession will be aligned with the surface normal direction of the perpendicular magnetic film. When the sample has a large magnetostriction constant $\lambda$, acoustic waves can tilt the magnetization and therefore create a detectable change in $M_z$ and will be captured as magnetization oscillations in the signal as well. Thus, the magnetization oscillation features captured by TR-MOKE contains both the spin precession and acoustic wave information (*20*). Therefore, analysing the data from both TDTR and TR-MOKE measurements allows for the investigation of the coupling between strain and spin. The TR-MOKE signals contain information of two behaviors, as the superposition of two frequencies in $M_z$: loosely categorized as the field-dependent one corresponding to the typical FM resonance resulting from coherent spin precession (ferromagnetic resonance, FMR), and the field-independent one caused by a time-dependent modulation of magnetic properties through strain.

In our TR-MOKE signals for all the [Co/Pd]$_n$ samples, the magnetization oscillations appear without $H_{ext}$, implying that these samples possess the magnetostriction effect, as discussed in Supplementary Note 1 and shown in Figs. S3 and S4. By fitting the TR-MOKE data, we find that in relatively thin films, the strain frequency (> 100 GHz) is much higher than that of spin precession (< 60 GHz) with a relatively low $H_{ext}$, as shown in Fig. S5. This frequency is for the lowest-frequency normal mode of the acoustic waves confined within the thin-film sample stack, whose half wavelength depends on the thickness of the entire sample stack (including the capping and seed layers). Therefore, by changing the sample thickness, the strain frequency can be tuned (*40*). In this study, we select the [Co(0.8 nm)/Pd(1.5 nm)]$_{11}$ multilayered structure with a relatively large (25 nm) thickness as a model system. This [Co(0.8 nm)/Pd(1.5 nm)]$_{11}$ multilayer is seeded with a Ta(5 nm)/Pd(5 nm) bilayer deposited on a SiO$_2$ (300 nm)/Si substrate and is capped with a 3-nm Ta layer. It has perpendicular anisotropy with an effective field of $H_{k,eff}$ ~ 6.5 kOe and magnetic anisotropy of $K_u$ ~ 4.4×10$^6$ erg/cm$^3$, as shown in Fig. 1B.

**Ultrafast measurement results.** The TR-MOKE results of [Co(0.8 nm)/Pd(1.5 nm)]$_{11}$ with the range of $H_{ext}$ from 10 kOe to 29 kOe are plotted in Fig. 2A. Interestingly, we find that for 18 kOe < $H_{ext}$ < 24 kOe, TR-MOKE signals show the amplitude of precessional oscillations of $M_z$ increases instead of the usual decrease in the first 60 ps following the pump excitation. However, when $H_{ext}$ is smaller than 18 kOe or larger than 24 kOe, the TR-MOKE signals in Fig. 2A appear as the typical monotonic decaying trend of damped oscillations (standard in TR-MOKE measurements of magnetization precession). Within the first several picoseconds, different energy carriers such as magnons, phonons, and electrons induced by the ultrafast laser pulse are out of equilibrium with each other. As the main goal of this work is to analyze the magnetic precession that is in the equilibrium regime, we purposefully start the fitting from 10 ps to avoid any possible non-equilibrium effect (indicated by the solid lines in Fig. 2A). From Fig. 2B of TDTR data, we can find that the acoustic strains prevail in the first 60 ps. For the range of 18 kOe < $H_{ext}$ < 24 kOe, the injected energy by the strain overcomes the magnetically dissipated energy, leading to the enhanced amplitude of spin precession roughly within this first 60 ps, in contrast to the monotonic decaying trend of a typical damped feature of spin precession at other fields. After 60 ps, the injected energy by the strain is insufficient to compensate for the dissipated energy due to the damping, and thus is not capable of maintaining the high energy state (the magneto-acoustic resonance state as explained later). The spin oscillation amplitude decreases, and the typical damped feature appears.

To further understand the magnetization dynamics and magneto-acoustic resonance resulting from the coupling between strain and spin, the TR-MOKE and TDTR signals of the [Co(0.8 nm)/Pd(1.5 nm)]$_{11}$ sample are analyzed in the time domain using the following equation:

$$V(t) = A + Be^{-t/C} + D_1 \sin(2\pi f_1 t + \varphi_1)e^{-t/\tau_1} + D_2 \sin(2\pi f_2 t + \varphi_2)e^{-t/\tau_2} \quad (1)$$

where $V$ is the signal, $t$ is the time, $A$, $B$, and $C$ are fitting parameters of the thermal background, $D_i$ represents the amplitude of sinusoid, $f_i$ is the frequency, $\varphi_i$ is the phase, and $\tau_i$ is the relaxation frequency. The indices of $i = 1, 2$ represent two separate modes (low-frequency Mode 1 and high-frequency Mode 2 as defined later), captured by the measurement simultaneously (*41,42*). The TDTR data only show one frequency corresponding to the picosecond acoustics (strain) in the system, which is independent of the external field. Fitting the TR-MOKE data (which captures the change in the *z*-component of magnetization) shows two distinct frequencies over most of the range: a frequency that depends on $H_{ext}$ following a Kittel dispersion (FMR), and a frequency that is independent of $H_{ext}$ that matches the frequency of the strain captured by TDTR (strain). More details about the amplitude and phase information of these two behaviors are provided in Supplementary Note 1. Two TR-MOKE signals, one in resonance ($H_{ext}$ = 21 kOe) and the other one out of magneto-acoustic resonance ($H_{ext}$ = 14 kOe), are chosen for comparison to study the coupling between spin precession and acoustic waves (Fig. 2C). We find that the frequency of the signal without the magneto-acoustic resonance shows two distinct peaks that are well separated, ~38 GHz for spin precession and ~60 GHz for the acoustic waves. However, for the signal at the magneto-acoustic resonance, two frequency peaks are overlapping at ~60 GHz, as plotted in Fig. 2D. This suggests that the magneto-acoustic resonance originates from the strong coupling between strain and spin.

**Extremely high-frequency magneto-acoustic resonance.** Figures 3A and 3C plot the frequencies of spin precession of the [Co(0.8 nm)/Pd(1.5 nm)]$_{11}$ multilayer, which are fitted from the TR-MOKE signal as a function of $H_{ext}$ based on Eq. (1). There are 2 modes, the low frequency mode being labeled Mode 1 (open black circles), and the high frequency mode being labeled Mode 2 (open red diamonds). These two modes are weakly coupled, *i.e.*, the coupling is small compared to the frequency difference, except near the resonance with a coupling coefficient $\kappa$ of ~ (16.5 GHz)$^2$ =272 GHz$^2$ that is determined by the magnetostriction coefficient, magnetic properties and the external field angle (details in Supplementary Note **3**D). At low fields where $\omega_1$ is much smaller than $\omega_2$, these modes are distinctly separated by the driving mechanism. The largest TR-MOKE signal corresponds to the low-frequency mode, driven directly by magnon from the fast demagnetization process. The frequency of Mode 1 ($\omega_1$) equals to the magnon frequency $\omega_M$ and is linearly dependent on $H_{ext}$, with a slope of $\gamma/2\pi$ with $\gamma$ being the gyromagnetic ratio. The displacement of this linear behavior of $\omega_1$ is caused by the effective anisotropy field. The amplitude $m_0$ depends on the initial precession angle (the maximum value ~1°), estimated based on the conversion from TR-MOKE voltage to the Kerr angle. $m_0$ decreases rapidly when increasing external magnetic field as $1/H_{ext}^2$, due to the energy gradient change according to the "law of approach to saturation" (*43*). The high-frequency Mode 2 is weakly detected because it is primarily phonon based. Its frequency $\omega_2$ is field independent at low fields, remaining at a constant 60 GHz, the phonon frequency $\omega_{ph}$ fitted from TDTR signals (blue stars). Its amplitude $\kappa\eta_0/(\omega_{ph}^2 - \omega_M^2)$ is proportional to the strain amplitude $\eta_0$ and the coupling coefficient $\kappa$ (Supplementary Note 3D). This indicates the high-frequency mode $\omega_2$ makes a negligible contribution to the magnon-dominated dynamics at low fields, verified by a slight side peak appearing in the FFT spectrum in Fig. 2C. The amplitude gradually increases when increasing the magnetic field, as the difference between $\omega_M$ and $\omega_{ph}$ shrinks.

When the frequencies of these two modes approach each other, high- and low-frequency modes hybridize, reflecting strongly coupled magnon-phonon dynamics. The two modes, driven by a quasiparticle with significant contributions from both magnon and phonon, deviate from their original characteristics discussed in the previous low-field condition. The frequencies of Modes 1 and 2 display an anti-crossing with a frequency gap $\Delta f$, deviating from the original $\omega_M$ ($\approx \omega_1$) or $\omega_{ph}$ ($\approx \omega_2$). The nature or presence of an anti-crossing feature cannot be proven to high confidence due to the experimental error bars, so our real evidence for anti-crossing comes from theory. Clearly, the strain drives the magnetization oscillations, which can only happen if they have the same symmetry. Therefore, there must be a mixing term in the Hamiltonian, and an anti-crossing is guaranteed. To quantitatively predict the gap $\Delta f$, we calculate the two frequencies of Modes 1 and 2 at the resonance point (defined as $\omega_M = \omega_{ph}$ at $H_{ext} \approx 21$ kOe), which gives $\omega_1 = \omega_{ph} - \left(\kappa/2\omega_{ph}\right)$ and $\omega_2 = \omega_{ph} + \left(\kappa/2\omega_{ph}\right)$. This yields $\Delta f = \omega_2 - \omega_1 = 4.6$ GHz, which is consistent with experimental data using no fitting parameters, as shown in Figs. 3A and 3C. The hybridization regime is defined as $H_{ext}$ at which $\left|\omega_M - \omega_{ph}\right|$ approximates $\kappa/\omega_{ph}$, leading to a range of 19.7 kOe < $H_{ext}$ < 22.1 kOe. The amplitudes of both modes increase in the hybridization regime, as shown in Figs. 3B and 3D. The original phonon-driven mode is now highly visible owing to the admixed magnon, where its amplitude changes from a negligible value (~ $\eta_0$) to a notable value. The original magnon-driven mode is enhanced owing to the pumping from phonon, where its amplitude enhancement is caused by the strain amplitude $\eta_0$, which also results in magnon-phonon hybridization. In addition, this hybridization appears as a magneto-acoustic resonance at the resonance field (the strongest effect) and nearby, as verified by the enhanced wave envelope in the TR-MOKE signals in Fig. 2A. We can conclude the magnon-phonon coupling substantially influences the spin dynamics and induces a resonant state. When further increasing $H_{ext}$ from the resonance-field range, the frequencies of these two modes move apart again. At such high fields, the hybridization disappears, and the modes are distinctly driven by the phonon or magnon. The low-frequency Mode 1 becomes the phonon-driven mode ($\omega_1 \approx \omega_{ph}$) while the high-frequency Mode 2 is the magnon-driven mode $\omega_2 = \omega_M$. Their behaviors are very similar to those at low fields.

**DISCUSSION**

To accurately describe the time-domain behavior and compare directly with TR-MOKE signals, we employ micromagnetic simulation based on the Landau-Lifthitz-Gilbert (LLG) equation, including the magnetostriction effect and the damping. The simulation reproduces the resonant state, whereby the wave envelope rises for ~100 ps and afterwards follows a standard relaxation, as shown in Fig. 4B. Comparing with the experimental data from TR-MOKE (Fig. 4A), we can see that the theoretical model we propose here can capture all the key features in this ultrafast magnon-phonon coupling behavior. From the energy perspective, the phonon initially pumps enough energy into the magnon system to overwhelm the energy dissipation (Fig. S6). This energy enhancement excites the magnetization to precess in a larger angle, presenting a rise, rather than the immediate decay found for the off-resonance condition, as observed in Fig. 2A for the low- and high-field cases. The simulation uses the same material magnetostriction coefficient (details in Supplementary Note 3), and thus the same coupling coefficient as used in the previous section to predict the frequencies of Modes 1 and 2. The rising time depends on the coupling coefficient $\kappa$. Relaxation occurs when the pumped energy by phonon is no longer sufficient to compensate the dissipated energy. The overall agreement of experiment and theory suggests that our assumption of a uniform spin mode is correct, although the existence of the lowest exchange magnon cannot be completely excluded (*29*).

**Time-dependent magnetization dynamics driven by acoustic waves.** To prove acoustic waves can manipulate the magnetization and further assist the switching on an ultrafast picosecond scale, we carried out the micromagnetic simulation of the out-of-plane magnetization ($M_z$) vs. the time delay with a square strain pulse. The strain pulse amplitude is 0.5%, the pulse period is 2 ns, and the pulse length is 1 ns. Figures 4D and 4E show $M_z$ vs. time delay of the system at two external fields near the magneto-acoustic resonance state. When the strain pulse is on, the system becomes excited rapidly to an enhanced large angle precession with a rise time of ~100 ps. When the strain pulse is off, the system at all two external fields shows relaxation behavior and the magnetization becomes aligned with $H_{ext}$. This large angle precession caused by the resonance between acoustic waves and spin precession is maintained steadily for non-decaying strain amplitude. In addition, as the spin dynamics is a nonlinear system, the resonance can happen in a wide field range of several kOe, centered at the resonance point with a field range determined by the coupling coefficient $\kappa$ (Fig. 4C). For example, an external magnetic field of $H_{ext}$ = 21 kOe has the maximum precession amplitude, while fields of $H_{ext}$ = 18 kOe or 24 kOe can also excite resonant behavior with relatively smaller precession amplitudes. It can also be seen that the precession amplitudes with 0.2% or 0.5% strain yield similar peak values. The magneto-acoustic resonance with $H_{ext}$ = 21 kOe is enhanced when the strain increases from 0.1% to 0.2% (see Fig. S7). This level of strain has been detected by previous researchers (*44,45*). Figure 4F presents the strain-assisted magnetization switching of the [Co(0.8 nm)/Pd(1.5 nm)]$_{11}$ system with $H_{ext}$ = 1 kOe. It is clearly seen that when no strain (strain amplitude of 0%) is present in the system, the magnetization of the system shows the normal damped decay, switching does not happen. When strain (strain amplitude of 0.5%) is applied, the system shows a large-angle magnetization precession and switching occurs in 3 ns. We also note that Vlasov *et al.* has recently predicted switching in an elliptical nanomagnet with shape anisotropy (*46*).

**CONCLUSION**

We experimentally detected acoustic waves with response time down to the order of 10 picoseconds in perpendicular magnetic [Co/Pd]$_n$ multilayers *via* a femtosecond laser pulse excitation. Through direct measurements of coherent phonons and magnetization, we observed a 60-GHz magneto-acoustic resonance when the frequencies of acoustic waves and spin precession approach each other. We developed a theoretical model and revealed the physical mechanism of magneto-acoustic resonance from the strain-spin interaction within an energy viewpoint. This model and experiment agree in their demonstration of hybridization between strain and spin waves near the resonance point. The results, by illustrating a pathway to switching the magnetization of a high anisotropy perpendicular material through strain-spin coupling, offer an approach to meeting the future technological needs of high speed and highly compact memory.

**MATERIALS AND METHODS**

**Sample preparation and characterization.** All samples with the stack of [Co($x$)/Pd($y$)]$_n$/Co($x$)/Ta(3) ($x$ = 0.30 ~ 0.70 nm; $y$ = 0.70 ~ 1.80 nm), from the bottom to top, are deposited on Si/SiO$_2$(300 nm) substrate at room temperature using a six-target Shamrock magnetron sputtering system with the ultrahigh vacuum (base pressure < 5.0 × 10$^{-8}$ Torr). The [Co(0.8 nm)/Pd(1.8 nm)]$_{11}$ multilayer is seeded with Ta(5 nm)/Pd(5 nm) bilayer, the others are seeded with Ta(3 nm)/Pd(3 nm) bilayer. All layers are sputtered with D.C. power sources and element targets under an Ar working pressure of 2.0 mTorr. The magnetic properties of all samples are characterized using a Physical Property Measurement System (PPMS) with the Vibrating Sample Magnetometer (VSM) module.

**TR-MOKE and TDTR measurements.** Both TDTR and TR-MOKE methods are based on the ultrafast pump-probe technique. In this technique, pump pulses are utilized to excite the sample to a higher energy level followed by a recovery process, while probe pulses detect this change as a function of time delay. A mode-locked Ti:Sapphire laser with a pulse duration of ~100 fs and a center wavelength of 783 nm at a repetition rate of 80 MHz is used for TDTR and TR-MOKE measurements. A 10× objective lens is used to produce a 1/$e^2$ spot radius of $w_0$ = 6 $\mu$m for both pump and probe beams. In addition, an external magnetic field of up to 29 kOe (at $\theta_H$ = 80° as defined in Fig. 1A) is applied in both TR-MOKE and TDTR measurements. TDTR signal is captured with a fast-response photodiode while the TR-MOKE signal is captured with a balanced detector. For our measurement setup, the field angle is limited to 80° ≤ $\theta_H$ ≤ 90° to achieve sufficiently high fields and to ensure a clear optical path for the laser beam. Therefore, we set $\theta_H$ to be 80° for all measurements of [Co/Pd]$_n$ samples, in order to achieve the highest precessional frequency with the smallest field and meanwhile to maximize the TR-MOKE signals (see Supplementary Note 1).

**Theory model and micromagnetic simulation.** In the theoretical model, we first derive the magnetic dynamics, in Eq. (2) under the influence of the magnetostriction and the damping effect, from the macrospin model. Unlike references *30* and *31*, the role of backaction is included, which allows predictions for the experimentally observed hybridization. We define that the $z'$-axis is titled from the original $z$-axis with an angle $\theta'_H$ ($\theta'_H = \pi/2 - \theta_H$) and perpendicular to the original $x$-axis. The equation for the out-of-plane magnetization $m_{z'}$ is:

$$\frac{d^2 m_{z'}}{dt^2} + \omega_M^2 m_{z'} + \Gamma \frac{dm_{z'}}{dt} = -\kappa \eta \tag{2}$$

where $\omega_M$ is the magnetic frequency, $\Gamma$ is the relaxation rate, and $\kappa$ is the coupling coefficient. The explicit expression for these parameters are:

$$\omega_M = \gamma \sqrt{\left(H_{ext} + H_{k0} \sin \theta'_H\right)\left(H_{ext} - H_{k0} \cos(2\theta'_H)\right)} \tag{3}$$

$$\Gamma = \alpha \gamma \left(2H_{ext} + H_{k0} \sin^2 \theta'_H - H_{k0} \cos(2\theta'_H)\right) \tag{4}$$

$$\kappa = 3\gamma^2 / 8 \left(H_{ext} + 2H_{k0} \sin^2 \theta'_H\right) \sin(2\theta'_H)(b_2/M_s) \tag{5}$$

where $\gamma$ is gyromagnetic ratio, $H_{ext}$ is external bias field, $H_{k0}$ is the effective perpendicular anisotropy, $b_2$ is the magnetostriction coefficient. Details may be found in Supplementary Note 3A.

The strain amplitude $\eta$ is governed by:

$$\frac{d^2 \eta}{dt^2} + \omega_{ph}^2 \eta = -\kappa m_{z'} \tag{6}$$

where $\omega_{ph}$ is the phonon frequency. Eq. (2) and Eq. (6) yields the characteristics of the two modes in the magnetic dynamics, as low frequency mode $\omega_1$ and high frequency mode $\omega_2$. Their amplitudes are $B_1$ and $B_2$ correspondingly. Approximations under varying conditions are performed in Supplementary Note 3D to aid physical understanding, and are summarized in Discussions.

$$\omega_1 = \sqrt{\left(\omega_M^2 + \omega_{ph}^2 - \sqrt{(\omega_M^2 - \omega_{ph}^2)^2 + 4\kappa^2}\right)/2} \tag{7}$$

$$\omega_2 = \sqrt{\left(\omega_M^2 + \omega_{ph}^2 + \sqrt{(\omega_M^2 - \omega_{ph}^2)^2 + 4\kappa^2}\right)/2} \tag{8}$$

$$B_1 = \frac{m_0 + \eta_0 \dfrac{\kappa}{\omega_M^2 - \omega_2^2}}{1 - \dfrac{\kappa}{\omega_M^2 - \omega_2^2} \dfrac{\kappa}{\omega_{ph}^2 - \omega_1^2}} \tag{9}$$

$$B_2 = \frac{-\dfrac{\kappa}{\omega_M^2 - \omega_2^2}\eta_0 - m_0 \dfrac{\kappa}{\omega_{ph}^2 - \omega_1^2}\dfrac{\kappa}{\omega_M^2 - \omega_2^2}}{1 - \dfrac{\kappa}{\omega_M^2 - \omega_2^2} \dfrac{\kappa}{\omega_{ph}^2 - \omega_1^2}} \tag{10}$$

where $m_0$ is the initial magnetization out-of-plane component $m_0$ deviating from the final equilibrium position. $\eta_0$ is the initial strain amplitude.

Thirdly, we perform micromagnetic simulation to accurately describe the magnetic dynamics versus time, including both the magnetostrictive and damping effect. We choose the experimental initial conditions in the simulation. Through a straightforward voltage to Kerr angle conversion in TR-MOKE signal, a precessional cone angle (maximum value ~ 1°) is used to set the initial magnetization condition in simulation (see Supplementary Note 2). A strain of 0.1% is used to set the initial strain condition.

**SUPPLEMENTARY MATERIALS**

Supplementary material for this article is available at http://advances.sciencemag.org/

**Acknowledgements:** We would like to thank Prof. Paul Crowell from University of Minnesota for valuable discussions and suggestions. **Funding:** This work was supported by C-SPIN, one of six centers of STARnet, a Semiconductor Research Corporation program, sponsored by MARCO and DARPA. D.M.L. and X.W. would like to thank the support from Advanced Storage Research Consortium (ASRC). **Author contributions:** D.L.Z., J.Z., T.Q., R.H.V, X.W., and J.P.W. conceived the research. D.L.Z. designed and prepared all of the samples and carried out all magnetic measurements. J.Z. and D.M.L. designed and carried out the TDTR and TR-MOKE measurements and fitted the data. D.L.Z. and J.Z. introduced the initial experimental results including magnetic properties and TDTR and TR-MOKE data on this topic to T.Q. and suggested the need for theoretical analysis. T.Q. carried out the theoretical prediction, analytical derivation, and micromagnetic simulation that inspired the experimental results of the resonance. D.L.Z and J.Z participated in the discussion of the theory and micromagnetic simulation. D.L.Z., J.Z., and T.Q. prepared the figures and drafted the manuscript. J.P.W., X.W., and R.H.V. coordinated the project. All the authors discussed the results and commented on the manuscript. **Competing interests:** The authors declare no competing interests. **Data and materials availability:** All data needed to evaluate the conclusions in the paper are present in the paper and/or the Supplementary Materials. Additional data related to this paper may be requested from the authors.

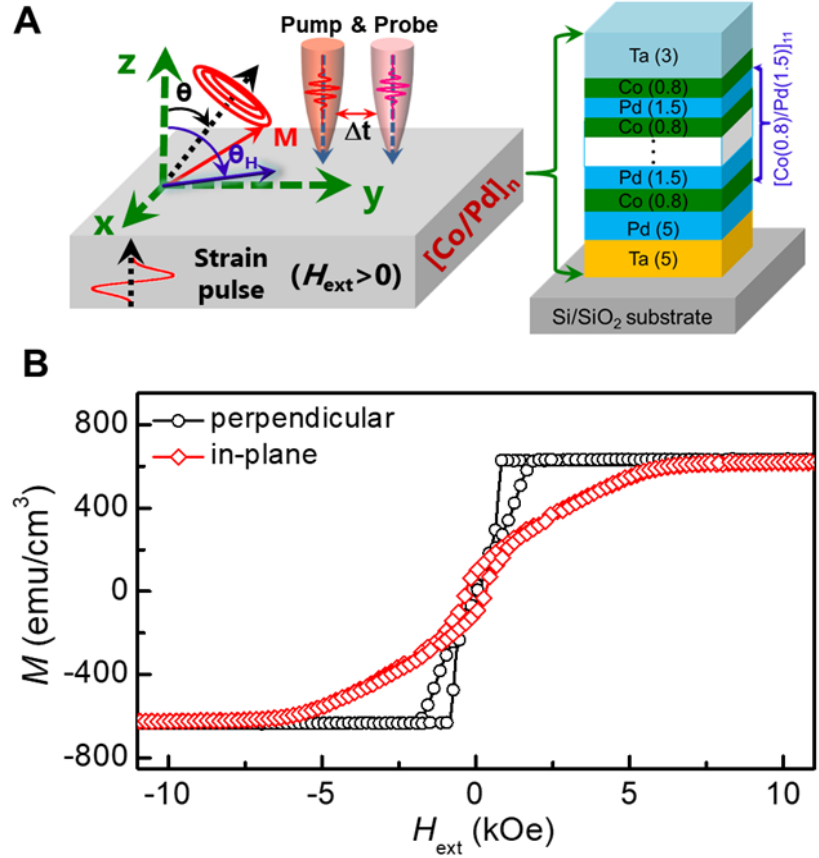

**Fig. 1. Measurement protocol.** (**A**) Illustration of the ultrafast time-resolved magneto-optical Kerr effect (TR-MOKE) measurements (left) on the [Co/Pd]$_n$ multilayer with numbers in parentheses denoting layer thicknesses in nanometres (right). In the TR-MOKE measurement, in the absence of an external magnetic field $H_{ext}$, the magnetostrictive effect can be measured, in which the acoustic strain wave induces the magnetization oscillation. The magnetization of [Co/Pd]$_n$ multilayer is tilted to the angle ($\theta$) when $H_{ext} > 0$ is applied with the angle ($\theta_H = 80°$). The TR-MOKE signals will include the signal from spin precession and acoustic strain wave. By fitting the data, we can separate them and identify their coupling. The figure in the right plane of the top shows the [Co/Pd]$_n$ multilayered structure used in our work. (**B**) The magnetic hysteresis (*M-H*) loops of the [Co(0.8 nm)/Pd(1.5 nm)]$_{11}$ multilayer with a magnetic anisotropy field $H_{k,eff}$ of ~6.5 kOe.

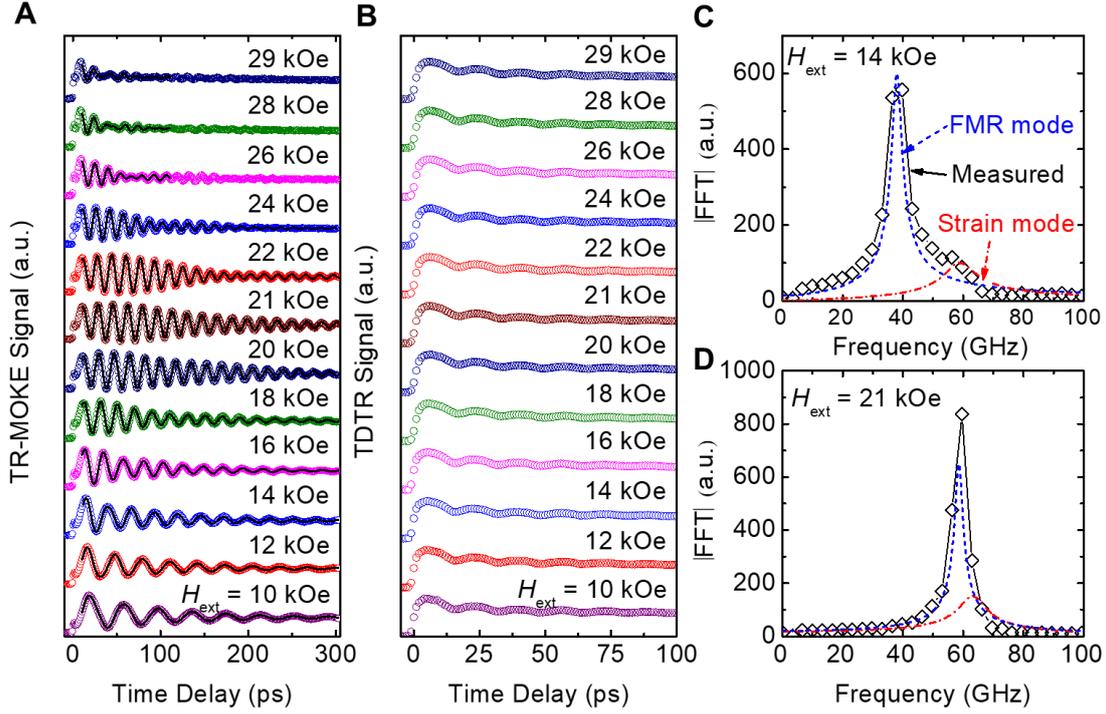

**Fig. 2. Ultrafast measurement results.** (**A**) The experimental and fitted TR-MOKE signals and (**B**) the experimental TDTR signals as a function of $H_{ext}$ (10 to 29 kOe). It is clearly seen that TDTR signals do not change in the whole region of $H_{ext}$ while TR-MOKE signals show different oscillation patterns with external fields. For $H_{ext}$ < 18 kOe or > 24 kOe, magnetization precession presents a damped oscillation, while for 18 kOe < $H_{ext}$ < 24 kOe, magnetization shows a resonance phenomenon. (**C**), (**D**) Fourier transform of the TR-MOKE signal with $H_{ext}$ = 14 kOe and 21 kOe, respectively, from which two peaks (FMR and strain) can be found. For $H_{ext}$ = 14 kOe, the two peaks are separate, however, the two peaks are overlapping when $H_{ext}$ = 21 kOe.

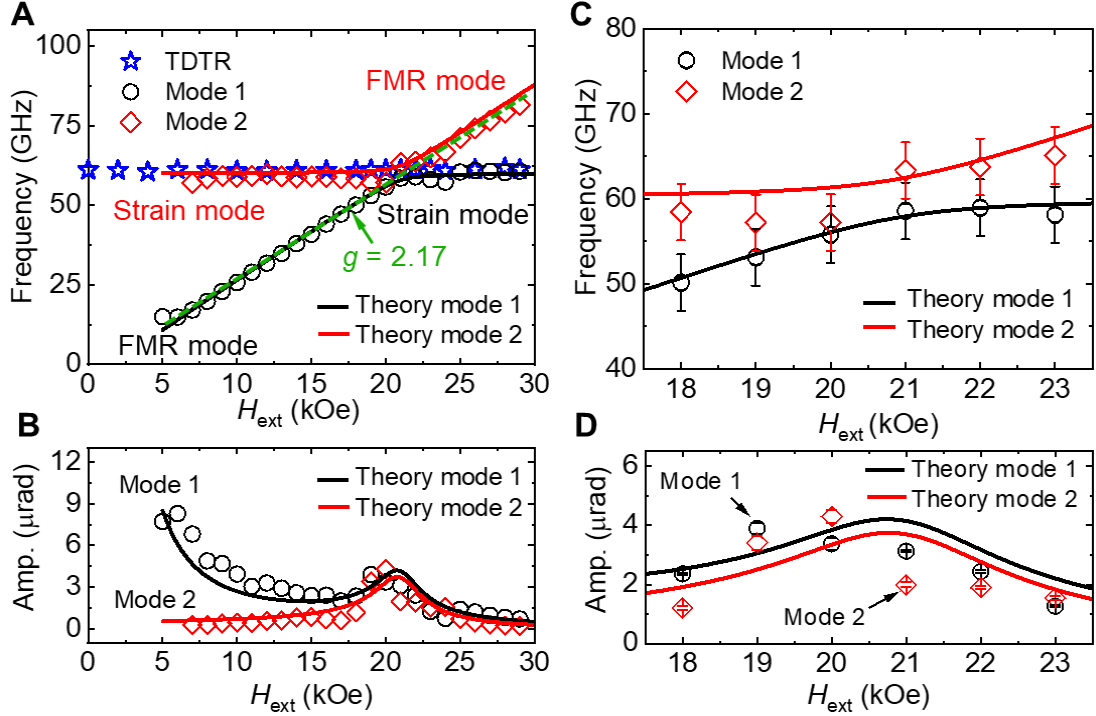

**Fig. 3. High-frequency magneto-acoustic resonance.** (**A**) The frequency of the [Co(0.8 nm)/Pd(1.5 nm)]$_{11}$ multilayer as a function of $H_{ext}$. Two frequencies of spin precession (Mode 1, open black circles and Mode 2, open red diamonds) are derived by fitting the experimental data of TR-MOKE. The figure also includes the frequency of acoustic waves measured from TDTR (blue stars). The anti-crossing point of Mode 1 and Mode 2 occurs at the resonance field ($H_{ext} \approx 21$ kOe), where the frequencies of Modes 1 and 2 split and open a gap $\Delta f$. We assign the strain mode as the one with field-independent frequencies that are nearly identical to the acoustic wave frequencies from TDTR. The frequency of the FMR mode increases linearly with $H_{ext}$. (**B**) The individual $M_z$ amplitudes of Modes 1 and 2 as a function of $H_{ext}$ for the [Co(0.8 nm)/Pd(1.5 nm)]$_{11}$ multilayer. There exists an apparent amplification of both modes due to the coupling between these two modes near the anti-crossing point. (**C**) A zoomed-in frequency plot of the resonance region. The error bars (~ 3 GHz) are based on the FFT resolution (~ 3 GHz). (**D**) A zoomed-in amplitude plot of the resonance region. The error bars represent the standard error from mathematical fitting of the measurement data based on Eq. (1).

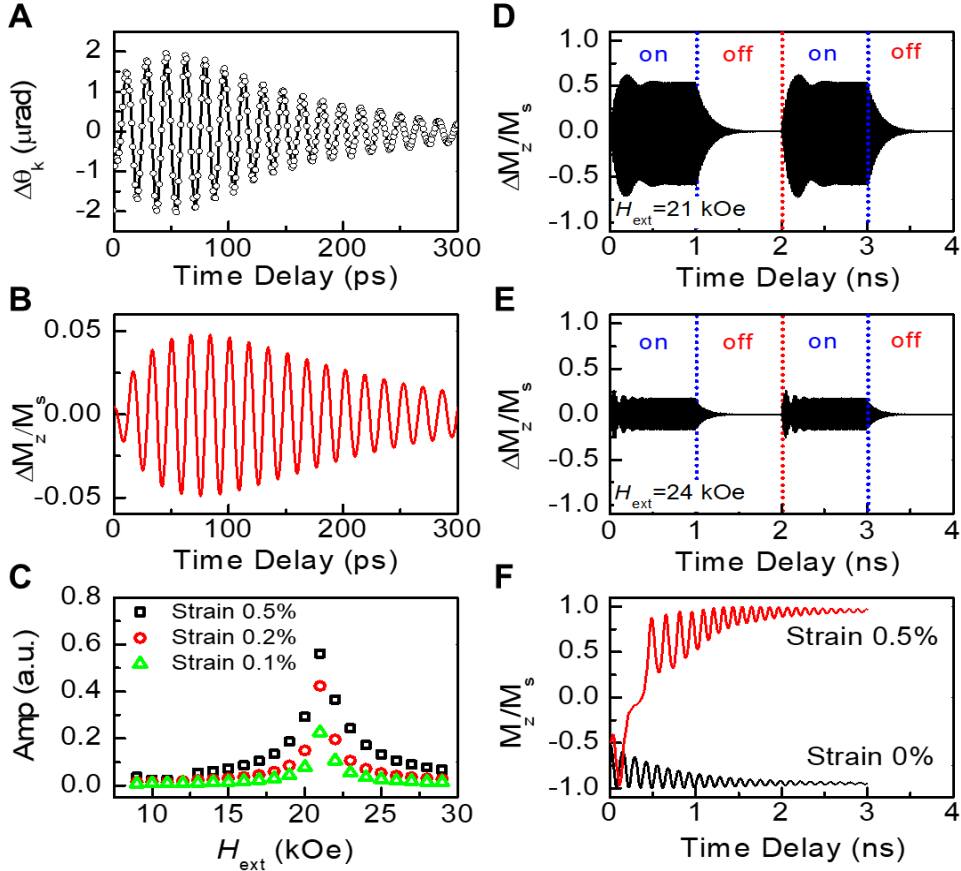

**Fig. 4. Time-dependent magnetization dynamics.** (**A**), (**B**) The experimental and simulated TR-MOKE signal of the [Co(0.8 nm)/Pd(1.5 nm)]$_{11}$ multilayer with $H_{ext}$ = 21 kOe, respectively. The strain used to produce the simulated signal is 0.5%. (**C**) The oscillation amplitude vs. $H_{ext}$ for different strain amplitudes (0.1%, 0.2% and 0.5%). (**D**), (**E**) The time evolution of the out-of-plane magnetization ($M_z$) vs. the time delay when a square strain pulse is applied. The pulse amplitude is 0.5%, the time period is 2.0 ns and the pulse length is 1.0 ns. The simulated spin precession coupled with a 0.5% strain pulse under $H_{ext}$ of 21 kOe (**D**) and 24 kOe (**E**), respectively. When the strain pulse is on, the system gets excited rapidly to an enhanced large angle precession with a rise time of ~100 ps under $H_{ext}$ = 21 kOe, but not 24 kOe. When the strain pulse is off, the system at these $H_{ext}$ values shows relaxation behavior. (**F**) Strain assisted switching for [Co(0.8 nm)/Pd(1.5 nm)]$_{11}$ with $H_{ext}$ = 1 kOe. When no strain (strain amplitude of 0%) is present in the system, switching does not occur, shown by the black line. When strain (strain amplitude of 0.5%) is applied, switching occurs, shown by the red line.

Supplementary Materials for

# High frequency magnetic resonance through strain-spin coupling in perpendicular magnetic multilayers

De-Lin Zhang[1†], Jie Zhu[2†], Tao Qu[3†], Dustin M. Lattery[2†], Randall H. Victora[1,3], Xiaojia Wang[1,2]* and Jian-Ping Wang[1]*

*Author to whom correspondence should be addressed: jpwang@umn.edu, wang4940@umn.edu and victora@umn.edu

**Supplementary Materials:**

Supplementary Note 1. Ultrafast TDTR and TR-MOKE Measurements

Supplementary Note 2. Voltage to Magnetization Conversion from TR-MOKE Signals

Supplementary Note 3. Micromagnetic Simulations

Figs. S1 to S9

**Supplementary Note 1. Ultrafast TDTR and TR-MOKE Measurements**

In this work, we investigate the coupling between strain and spin and magneto-acoustic resonance by using TR-MOKE and TDTR methods. The mode-locked Ti:Sapphire laser, with a duration of ~ 100 fs and a center wavelength of 783 nm at a repetition rate of 80 MHz is employed to capture the TDTR signals. A beam splitter separates the laser into pump and probe beams with orthogonal polarizations. An electro-optical modulator (EOM) synchronized to a function generator modulates the pump beam at 9 MHz, while a mechanical chopper modulates the probe sensing at 200 Hz. A relative optical path between the pump beam and probe beam is adjusted via a mechanical delay stage, which produces a time separation between pump excitation and probe sensing ranging from −100 ps to 4 ns. A 10× objective lens is used to produce a $1/e^2$ spot radius of $w_0 = 6$ $\mu$m for both pump and probe beams. The thermal response of a sample subjected to a modulated pump heating was characterized by measuring the temperature induced intensity variation of a probe beam reflected from the sample surface.

TR-MOKE is an upgraded version of the standard TDTR system. In this work, the polar-MOKE configuration is applied, where the Kerr rotation angle is proportional to the change of the out-of-plane component of magnetization ($M_z$). The major difference in TR-MOKE is that the change in polarization state of the probe beam is detected instead of the reflectance in TDTR. The reflected probe beam is split into two paths with orthogonal polarization states by a Wollaston prism. The changes in the relative intensities of those two paths are detected with a balanced detector using a differential channel. This differential channel monitors the transient change in the polarization state of the probe beam reflected from the sample, which is related to the Kerr rotation angle (*39,47-50*). In addition, an external magnetic field ($H_{ext}$) up to 29 kOe is applied at an angle from the surface-normal $\theta_H$ in both TR-MOKE and TDTR measurements to tilt the equilibrium magnetization direction of the films. Usually, the stimulation of the films from laser pulse generates not only the temperature rise, but also acoustic strain waves (ASWs) as a result of instantaneous thermal expansion. In TDTR, ASWs are captured by collecting the picosecond acoustic signals. In TR-MOKE, for ferromagnetic films with large magnetostriction coefficient, the magnetization is strongly coupled with ASWs, and thus the TR-MOKE signals will contain the features from magnetization precession and acoustic vibration. The schematics of TDTR and additional accessories for upgraded TR-MOKE are shown in Fig. S1.

Typical TR-MOKE measurement signals will contain a small amount of non-MOKE components, which can be removed by subtracting the TR-MOKE signals obtained with two opposite magnetic directions. To show that this non-MOKE component contributes negligibly to the magnetization precession and acoustic waves, we conduct measurements of the optimal structure, the [Co(0.8 nm)/Pd(1.5 nm)]$_{11}$ multilayer, at the resonance field of $H_{\text{ext}} \approx 21$ kOe. By magnetizing the sample with a positive field, the magnetization of sample will be in the positive *y-z* plane (see the definition of coordinates in Fig. 1A of the main text). The corresponding TR-MOKE signal is set to be positive and labeled as $M^+$. When the field direction is reversed, the sample magnetization is along the opposite direction (the negative *y-z* plane). The resulting TR-MOKE signal is set to be negative and labeled as $M^-$. The subtraction of these two measurements ($M^+ - M^-$) can remove the non-MOKE components of TR-MOKE signals (*49,50*). Fig. S2A shows the raw TR-MOKE signals of $M^+$, $M^-$, and the subtracted signal of $M^+ - M^-$ for the [Co/Pd]$_n$ multilayer near resonance. The fact that both the strain behavior and FMR behavior are maintained in the subtracted signal of $M^+ - M^-$ (and nearly double the amplitudes) shows that both behaviors are resulting from the MOKE response of the sample magnetization. The resonance frequencies and relaxation times of the strain behavior and FMR behavior, extracted from the raw TR-MOKE signals of $M^+$, $M^-$, and the subtracted signal of $M^+ - M^-$, are depicted in Figs. S2B and S2C, respectively. The differences in the values analyzed using these three sets of TR-MOKE signals are less than 1% for the resonance frequency and less than 5% for the relaxation time, suggesting a negligible contribution from the non-MOKE components in TR-MOKE signals. Thus, for future data analysis, we choose to show only the $M^+$ signals.

To demonstrate the magnetostriction effect of [Co (0.4 nm)/Pd (0.9 nm)]$_n$ multilayers, we prepared these [Co/Pd]$_n$ multilayers by changing the repeat period ($n$ = 3, 5, 7, 9). They all exhibit an out-of-plane easy axis (perpendicular magnetic anisotropy). The TR-MOKE and TDTR data measured without $H_{ext}$ are shown in Fig. S3. We can clearly see the apparent oscillation as a function of the time separation between the pump heating and probe sensing, suggesting that these [Co/Pd]$_n$ multilayers possess the magnetostriction effect excited by femtosecond laser pulse. Similar TR-MOKE results are observed in the [Co($x$)/Pd($y$)]$_7$ multilayers with $x$ = 0.3~0.7 nm; $y$ = 0.9 nm and $x$ = 0.4 nm; $y$ = 0.7~1.8 nm, as shown in Fig. S4. The frequency of ASWs is fitted to be above 100 GHz. Figures S5A and S5B show the experimental and fitting TR-MOKE and TDTR signals of the [Co(0.7 nm)/Pd(1.5 nm)]$_7$ multilayer with in-plane $H_{ext}$ (0 - 20 kOe). Oscillation features are observed in both TDTR and TR-MOKE measurements. The TR-MOKE signal measured at $H_{ext}$ = 3 kOe is converted to frequency domain via the Fourier transform as shown in Fig. S5C, two peaks (34 GHz of spin precession and 107 GHz of ASWs) can be found simultaneously. The individual frequencies of two behaviors (corresponding to FMR and strain behaviors) extracted from fitting TR-MOKE signals are plotted as a function of $H_{ext}$ in Fig. S5D. We found the frequency of the strain behavior remains at a constant value of 107 GHz; however, the frequency of FMR behavior first decreases from 45 GHz ($H_{ext}$ = 3 kOe) to 18 GHz ($H_{ext} \approx H_{k,eff} \approx$15 kOe) and then increases up to 30 GHz at $H_{ext}$ of 20 kOe. It is apparent that the frequency of spin precession increases linearly with $H_{ext}$ when $H_{ext}$ is much higher than $H_{k,eff}$. Thus, we anticipate that higher than 100 GHz magneto-acoustic resonance can be produced with a thinner sample, a larger $H_{ext,}$ or with a smaller anisotropy.

**Supplementary Note 2. Voltage to Magnetization Conversion from TR-MOKE Signals**

In order to accurately model the magnetization dynamics in the system with micromagnetic simulations, it is important to choose the correct initial condition for the magnetization. This is usually done by choosing an initial condition with a magnetization angle away from equilibrium (an initial cone angle) in the simulation that allows the magnetization to precess. The TR-MOKE signal can be converted to a precessional cone angle through a straightforward three-step conversion. First, the measured TR-MOKE signal (in μV) is converted to a Kerr angle (in degrees) *via* the detector linear responsivity, based on which, 1 V of TR-MOKE signal in the differential channel of the balanced detector corresponds to 10° of Kerr rotation ($\theta_k$). Next, the change in $\theta_k$ with respect to the change in magnetization can be related to the saturation properties *via* $d\theta_k/dM_z \approx \theta_{ks}/M_s$. Using the literature value of $\theta_{ks} = 0.15°$ for [Co/Pd]$_n$ (*49*), $M_s$ from VSM measurements, and $d\theta_k$ converted from TR-MOKE signals, we can find $dM_z$, the change in $M_z$. Finally, the cone angle is calculated based on the amplitude of $dM_z$ and the equilibrium direction of magnetization ($\theta$). A typical TR-MOKE measurement with a few tens of μV in signal will lead to a cone angle of less than 1°. For this reason, we choose a cone angle of 1° as the initial condition for the magnetization to precess in our the micromagnetic simulations. With this initial cone angle of 1°, we can achieve magneto-acoustic resonance with a strain of ~0.1%. This 0.1% amplitude of strain is achievable for ultrafast-laser based dynamic measurements, as a result of thermal expansion launched by laser heating. Considering the relatively small-amplitude strain (0.1%) from optical excitation compared with mechanical strains, we do not anticipate any hysteresis effect to occur in the sample system during TR-MOKE measurements. In addition, we employ volume conservation in simulation, which does not include the hysteresis effect either.

**Supplementary Note 3. Micromagnetic Simulations**

**A. Temporal strain effective perpendicular anisotropy**

In order to simulate the coupling between ASWs and spin precession, we introduce a magnetostrictive energy term $F_{me}(\mathbf{r},t)$ as a source of the ASWs in the free energy density for the [Co/Pd]$_n$ multilayer systems.

$$\begin{aligned}F_{me}(\mathbf{r},t) = & b_1(m_{x'}(\mathbf{r},t)^2 e_{x'x'} + m_{y'}(\mathbf{r},t)^2 e_{y'y'} + m_{z'}(\mathbf{r},t)^2 e_{z'z'}) \\ & + b_2(m_{x'}(\mathbf{r},t)m_{y'}(\mathbf{r},t)e_{x'y'} + m_{x'}(\mathbf{r},t)m_{z'}(\mathbf{r},t)e_{x'z'} + m_{y'}(\mathbf{r},t)m_{z'}(\mathbf{r},t)e_{y'z'})\end{aligned}$$ (S1)

The magnetostriction coefficients $b_1 = -16\times10^7$ erg/cm$^3$ and $b_2 = 26\times10^7$ erg/cm$^3$ for face-centered-cubic Co (*51*). $x', y', z'$ represent the crystallographic Cartesian coordinate, in which, the $z'$ axis is along the [001] crystallographic orientation with respect to the coordinate origin. The other coordinate we use is the thin film coordinate, represented by $x, y, z$ such that the $z$-axis is perpendicular to the thin film, corresponding to the crystallographic [111] orientation. $\mathbf{m}_{cryst} = m_i(\mathbf{r})(i = x', y', z')$ is the projected component of magnetization vector $\mathbf{m}$ at position $\mathbf{r}$ to the crystallographic coordinates. Correspondingly, $\mathbf{m}_{lab} = m_i(\mathbf{r})(i = x, y, z)$ is the projected component to the thin-film coordinate. Position $\mathbf{r}$ is always marked in the thin-film coordinate. We only consider the uniformly applied strain tensor $\ddot{e}$ such that the whole observable film is under compression/expansion unitarily. Thus the spatial dependence of the strain tensor is not involved. In the thin film coordinate, we define longitudinal amplitude components to be in the perpendicular direction of the thin film, while the transverse components are in the in-plane direction of the thin film. The amplitudes of these two components obey volume conservation. We can estimate their amplitude using first order approximation, because the ASWs are a small perturbation to the system. In the thin film coordinate, the strain tensor $\ddot{e}_{lab}$ is assumed isotropic with a Poisson's ratio of 0.33.

$$\ddot{e}_{lab} = \begin{pmatrix} -\dfrac{\eta(t)}{2} & 0 & 0 \\ 0 & -\dfrac{\eta(t)}{2} & 0 \\ 0 & 0 & \eta(t) \end{pmatrix} \qquad (S2)$$

$\eta(t)$ is the time-dependent acoustic strain, written as $\eta(t) = \eta_0 \cos(2\pi f_s t)\exp(-t/\tau)$, where $\eta_0$ is the strain amplitude (~0.1%), $f_s$ and $\tau$ are the strain frequency (tuned by the [Co/Pd]$_n$ multilayer) and strain relaxation time, respectively.

To couple the strain tensor and the magnetization, we transform the strain tensor from the thin film coordinate to the crystallographic coordinate applying a rotation matrix $\mathfrak{R}$. In the crystallographic coordinate,

$$\Re = \begin{pmatrix} -\sin\phi & \cos\phi & 0 \\ -\cos\theta\cos\phi & -\cos\theta\sin\phi & \sin\theta \\ \sin\theta\cos\phi & \sin\theta\sin\phi & \cos\theta \end{pmatrix} \quad \text{(S3)}$$

$$\vec{e}_{cryst} = \Re^T \vec{e}_{lab} \Re = \begin{pmatrix} 0 & \frac{\eta(t)}{2} & \frac{\eta(t)}{2} \\ \frac{\eta(t)}{2} & 0 & \frac{\eta(t)}{2} \\ \frac{\eta(t)}{2} & \frac{\eta(t)}{2} & 0 \end{pmatrix} \quad \text{(S4)}$$

$$\mathbf{m}_{cryst} = \Re^T \mathbf{m}_{lab} \quad \text{(S5)}$$

where, $\theta = \cos(1/\sqrt{3})$ and $\phi = \frac{\pi}{4}$, according to the crystallographic [111] orientation in the thin film coordinate.

Substituting $\vec{e}_{cryst}$ and $\mathbf{m}_{cryst}$ into Eq. (S1), $F_{me}$ can be simplified as

$$F_{me} = (3/4) b_2 \eta(t) m_Z^2 \quad \text{(S6)}$$

Thus, the magnetostriction energy can be viewed as an effective perpendicular anisotropy, with time-varying behavior. For the strain amplitude of 0.1%, the amplitude of the effective strain anisotropy is $1.95 \times 10^5$ erg/cm$^3$, smaller than the effective crystalline uniaxial anisotropy $1.5 \times 10^6$ erg/cm$^3$ at room temperature (*51*). Its temporal dependence will introduce a magneto-acoustic resonance phenomenon in the spintronic system.

## B. Landau–Lifshitz–Gilbert (LLG) equation

After obtaining the temporal strain effective perpendicular anisotropy $F_{me}(\mathbf{r},t)$, the phenomenological expression of the total free energy density $F(\mathbf{M},\mathbf{r}) = F_{ani} + F_{me} + F_d + F_{ex} + F_{zeeman}$ for [Co/Pd]$_n$ multilayers with <111> crystallographic orientation deposited on Si/SiO$_2$ (300 nm) substrate is

$$F_{ani}(\mathbf{r},t) = K_{uni}(1 - M_z(\mathbf{r})^2) \quad \text{(S7)}$$

$$F_{me}(\mathbf{r},t) = \frac{3}{4} b_2 \eta(t) m_z^2 \quad \text{(S8)}$$

$$F_\text{d}(\mathbf{r},\text{t}) = \mathbf{M}(\mathbf{r})\int d\mathbf{r}' \ddot{\tilde{N}}(\mathbf{r}-\mathbf{r}')\mathbf{M}(\mathbf{r}') \tag{S9}$$

$$F_\text{ex}(\mathbf{r},\text{t}) = A(\nabla \mathbf{M}(\mathbf{r}))^2 \tag{S10}$$

$$F_\text{zeeman}(\mathbf{r},\text{t}) = \mathbf{H}\cdot\mathbf{M}(\mathbf{r}) \tag{S11}$$

where $F_\text{ani}$, $F_\text{d}$, $F_\text{ex}$ and $F_\text{zeeman}$ represent the magnetocrystalline uniaxial anisotropy, magnetostatic energy, exchange energy, and Zeeman energy, respectively. $M_\text{s}$ and $K_\text{uni}$ are the saturation magnetization and uniaxial anisotropy, which can be extracted from the magnetic hysteresis (*M-H*) loops. A is the exchange constant; **H** is the external magnetic field. $\ddot{\tilde{N}}(\mathbf{r}-\mathbf{r}')$ is the magnetostatic tensor with spatial dependence.

The spin dynamics with magnetostriction included, are described by the LLG equation,

$$\frac{d\mathbf{M}}{dt} = -\frac{\gamma}{1+\alpha^2}\left[\mathbf{M}\times\mathbf{H}_\text{eff} + \frac{\alpha}{M_\text{S}}\mathbf{M}\times(\mathbf{M}\times\mathbf{H}_\text{eff})\right] \tag{S12}$$

where the first term describes the torque driving the precession of the magnetization vector around the effective time-dependent magnetic field $\mathbf{H}_\text{eff}(\mathbf{r},\text{t})$,

$$\mathbf{H}_\text{eff}(\mathbf{r},\text{t}) = -\frac{\partial F(\mathbf{r},\text{t})}{\partial \mathbf{M}(\mathbf{r},\text{t})} \tag{S13}$$

The second term describes precession damping according to the phenomenological Gilbert damping parameter $\alpha = 0.02$ (*52*), $\gamma$ is the gyromagnetic ratio.

## C. Numerical simulation for magneto-acoustic resonance

During the simulation, we track the energy density change versus time at different strain amplitudes and compare it with the dynamics of the out-of-plane magnetization component; this helps us gain an insight into how the strain functions to enhance the TR-MOKE signal at the magneto-acoustic resonance state observed in experiments, shown in Fig. S6. The energy density is the spatial average of the free energy $F(\mathbf{r},\text{t})$ value in the spin system. When the ASWs are absent in Fig. S6A, the energy dissipates due to the damping and the energy density keeps decreasing. The $M_\text{z}$ manifests a damped motion and the amplitude shows an exponential relaxation, in Fig. S6B. When the strain is present, the ASWs can pump energy into the spin system and influence the spin dynamics. When the strain amplitude is 0.005%, during the beginning 100 ps, the system is near a threshold such that the pumped energy can compensate the dissipated energy, shown in Fig. S6C. The total energy shows oscillations caused by the strain. The spin dynamics shows a constant amplitude, in Fig. S6D, that is neither relaxation nor resonance behavior. After the beginning 100 ps, the strain decays into an amplitude much smaller than its initial value (0.005%). The pumped energy can be neglected compared with the energy dissipation, with the total energy behaving similarly to the case in the absence of strain. Accordingly, the spins decay into the equilibrium axis. When the strain amplitude is 0.1%, for the initial 100 ps, the pumped energy dominates and wins over the dissipated energy (Fig. S6E), thus the spin system can be excited to a high energy state. This excited state represents the magneto-acoustic resonance and can be maintained if a way to produce stable ASWs is applied (Fig. S6F). Afterwards, as the amplitude of the ASWs decays, the pumped energy is insufficient to equilibrate the dissipated energy and the free energy decreases, thus the system relaxes to a minimum energy state, the same as the final state in 0% strain case.

A large-angle magnetization precession can be initiated through the ASWs pulses, shown in Fig. S7. Through the coupling between the strain and spin, the energy transferred from strain can be an efficient contribution to excite the magnetization oscillation. The magnetization responds quickly to the strain, proceeding from the stationary equilibrium state to the large oscillation around this axis. The rising time is ~200 ps. When the strain pulse is off, the magnetization relaxes to its equilibrium state. The oscillation also shows a nonlinear property, that a relatively small strain can still achieve a large oscillation amplitude. For example, identifying the oscillation amplitude through the oscillation cone angle, the angle of 0.1% strain yields a 12.5° cone angle, with the change of the magnetization $\Delta M_z/M_s$ ~40%.

### D. Analytic understanding of the coupled ASW-spin systems

To analytically interpret the function of ASWs, we can average the magnetization $\mathbf{M}(\mathbf{r},t)$ spatially in the microspin LLG equation and rewrite it, (as the TR-MOKE experiment is measuring the average magnetization in a micrometer spot). $x', y', z'$ are the field coordinate in which the $y'$ axis is the bias field direction, as shown in Fig. S8.

$$\frac{d\mathbf{m}}{dt} = -\gamma(\mathbf{m} \times \mathbf{H}_{eff}) + \alpha \mathbf{m} \times \frac{d\mathbf{m}}{dt} \tag{S14}$$

$$\frac{dm_x}{dt} = -\gamma(m_y H_{eff,z} - m_z H_{eff,y}) + \alpha(m_y \frac{dm_z}{dt} - m_z \frac{dm_y}{dt}) \tag{S15}$$

$$\frac{dm_y}{dt} = -\gamma(-m_x H_{eff,z} + m_z H_{eff,x}) + \alpha(-m_x \frac{dm_z}{dt} + m_z \frac{dm_x}{dt}) \tag{S16}$$

$$H_{eff,x} = 0 \tag{S17}$$

$$H_{eff,y} = H_{ext} \cos\theta'_H \tag{S18}$$

$$H_{eff,z} = -4\pi M_s m_z + 2(K_{uni} + \delta K \sin(\omega_0 t))/M_s m_z + H_{ext} \sin\theta'_H \tag{S19}$$

where $\mathbf{m}$ is the normalized unit vector of the average magnetization and $\theta'_H = \pi/2 - \theta_H$.

The rotation matrix $\tilde{R}$ is used to rotate the film coordinate to the new coordinate that sets the field direction as the new $z$ axis.

$$\tilde{R} = \begin{pmatrix} 1 & 0 & 0 \\ 0 & \cos(\theta'_H) & -\sin(\theta'_H) \\ 0 & \sin(\theta'_H) & \cos(\theta'_H) \end{pmatrix} \tag{S20}$$

Transforming both the effective field $\mathbf{H}_{eff}$ and the unit vector of magnetization $\mathbf{m}$ between the film coordinate and the field coordinate, we can get:

$$\mathbf{H}'_{eff} = \tilde{R}\mathbf{H}_{eff} \tag{S21}$$

$$\mathbf{m} = \tilde{R}^{-1}\mathbf{m}' \tag{S22}$$

Thus the $\mathbf{H}'_{eff}$ can be rewritten as

$$\mathbf{H}'_{eff} = \begin{Bmatrix} 0 \\ H_{ext} + 2\frac{K_{eff} + \delta K \sin(\omega_0 t)}{M_s}(-\cos\theta'_H m_{z'} + \sin\theta'_H m_{y'})\sin\theta'_H \\ -2\frac{K_{eff} + \delta K \sin(\omega_0 t)}{M_s}(-\cos\theta'_H m_{z'} + \sin\theta'_H m_{y'})\cos\theta'_H \end{Bmatrix} \tag{S23}$$

where the effective anisotropy $K_{eff} = K_{uni} - 2\pi M_s^2$. As the precession amplitudes of the magnetization ($\delta m_{x'}$ and $\delta m_{z'}$) is small, within the first order approximation, we can set, $\delta m_{x'} \ll 1$, $\delta m_{z'} \ll 1$, and $\langle m_{y'} \rangle \approx 1$. Thus, $\mathbf{H}_{\text{eff}}$ is simplified as

$$\mathbf{H}'_{\text{eff}} = \begin{pmatrix} 0 \\ H_{\text{ext}} + H_k(-\cos\theta'_H m_{z'} + \sin\theta'_H)\sin\theta'_H \\ -H_k(-\cos\theta'_H m_{z'} + \sin\theta'_H)\cos\theta'_H \end{pmatrix} \quad (S24)$$

where $H_k = 2(K_{\text{eff}} + \delta K \sin(\omega_0 t))/M_s$ with $\delta K$ being the amplitude of the strain effective perpendicular anisotropy. Further, we define $H_{k0} = 2K_{\text{eff}}/M_s$ to simplify the following derivation. Considering $dm_{y'}/dt \approx 0$, we can get the second-order derivatives of $m_{x'}$ and $m_{z'}$ when only keeping the first order:

$$\frac{d^2 m_{z'}}{dt^2} + \omega_M^2 m_{z'} + \Gamma \frac{dm_{z'}}{dt} = -\kappa\eta \quad (S25)$$

where $\omega_M$ is the magnetic frequency, $\Gamma$ is the relaxation rate, and $\kappa$ is the coupling coefficient. The explicit expressions for these parameters are:

$$\omega_M = \gamma\sqrt{(H_{\text{ext}} + H_{k0}\sin\theta'_H)(H_{\text{ext}} - H_{k0}\cos(2\theta'_H))} \quad (S26)$$

$$\Gamma = \alpha\gamma\left(2H_{\text{ext}} + H_{k0}\sin^2\theta'_H - H_{k0}\cos(2\theta'_H)\right) \quad (S27)$$

$$\kappa = 3\gamma^2/8\left(H_{\text{ext}} + 2H_{k0}\sin^2\theta'_H\right)\sin(2\theta'_H)(b_2/M_S) \quad (S28)$$

where $\gamma$ is gyromagnetic ratio, $H_{\text{ext}}$ is external magnetic field, $H_{k0}$ is the effective perpendicular anisotropy, $\theta'_H = \pi/2 - \theta_H$ with $\theta_H$ being the angle between external magnetic field and the surface normal as defined in Fig. **1**, $b_2$ is the magnetostriction coefficient. From the coupling term, the magnetic oscillation amplitude is related to the angle $\theta'_H$ and the strain amplitude. Ideally, with the increase of $\theta'_H$ or the strain amplitude, we should be able to enhance the oscillation amplitude and gain a more significant magneto-acoustic resonance.

The equations for two coupled harmonic oscillators are:

$$\frac{d^2 m_{z'}}{dt^2} + \omega_M^2 m_{z'} = -\kappa\eta \quad (S29)$$

$$\frac{d^2\eta}{dt^2} + \omega_{ph}^2 \eta = -\kappa m_{z'} \quad (S30)$$

where the $m_{z'}$ is the out-of-plane component of magnetization, $\eta$ is the strain induced by the laser heating, $\omega_M$ is the Kittle frequency, and $\omega_{ph}$ is the frequency of the lowest order normal phonon mode.

Setting the determinant to zero yields:

$$\omega_1 = \frac{\sqrt{\omega_M^2 + \omega_{ph}^2 - \sqrt{(\omega_M^2 - \omega_{ph}^2)^2 + 4\kappa^2}}}{\sqrt{2}} \quad (S31)$$

$$\omega_2 = \frac{\sqrt{\omega_M^2 + \omega_{ph}^2 + \sqrt{(\omega_M^2 - \omega_{ph}^2)^2 + 4\kappa^2}}}{\sqrt{2}} \quad (S32)$$

The general solutions are:

$$m_{z'}(t) = B_{11}^+ \cos\omega_1 t + B_{12}^+ \cos\omega_2 t + B_{11}^- \sin\omega_1 t + B_{12}^- \sin\omega_2 t \quad (S33)$$

$$\eta(t) = B_{21}^+ \cos\omega_1 t + B_{22}^+ \cos\omega_2 t + B_{21}^- \sin\omega_1 t + B_{22}^- \sin\omega_2 t \quad (S34)$$

The amplitudes $B_{ij}^+$ and $B_{ij}^-$ are not all independent: plugging the solutions back to Eqs. (S29) and (S30) implies:

$$\text{For } \omega = \omega_1,\ B_{21}^+ = -\frac{\kappa}{\omega_{ph}^2 - \omega_1^2} B_{11}^+,\ B_{21}^- = -\frac{\kappa}{\omega_{ph}^2 - \omega_1^2} B_{11}^- \quad (S35)$$

$$\text{For } \omega = \omega_2,\ B_{12}^+ = -\frac{\kappa}{\omega_M^2 - \omega_2^2} B_{22}^+,\ B_{12}^- = -\frac{\kappa}{\omega_M^2 - \omega_2^2} B_{22}^- \quad (S36)$$

The initial conditions are the time derivative and the initial values of the magnetization and the strain at t=0

$$m_{z'}(0) = m_0 \quad (S37)$$

$$\dot{m}_{z'}(0) = 0 \quad (S38)$$

$$\eta(0) = \eta_0 \quad (S39)$$

$$\dot{\eta}(0) = 0 \quad (S40)$$

The initial conditions remove sin terms. The solutions are simplified as:

$$m_{z'}(t) = -\frac{\kappa}{\omega_M^2 - \omega_2^2} B_2 \cos \omega_1 t + B_1 \cos \omega_2 t \tag{S41}$$

$$\eta(t) = B_2 \cos \omega_1 t - \frac{\kappa}{\omega_{ph}^2 - \omega_1^2} B_1 \cos \omega_2 t \tag{S42}$$

where $B_{11}^+$ in Eq. (S33) is written as $B_1$, and $B_{22}^+$ in Eq. (S34) is written as $B_2$. This yields:

$$B_1 - \frac{\kappa}{\omega_M^2 - \omega_2^2} B_2 = m_0 \tag{S43}$$

$$B_2 - \frac{\kappa}{\omega_{ph}^2 - \omega_1^2} B_1 = \eta_0 \tag{S44}$$

Solving the coupled equations yields:

$$B_1 = \frac{m_0 + \eta_0 \frac{\kappa}{\omega_M^2 - \omega_2^2}}{1 - \frac{\kappa}{\omega_M^2 - \omega_2^2} \frac{\kappa}{\omega_{ph}^2 - \omega_1^2}} \tag{S45}$$

$$B_2 = \frac{\eta_0 + m_0 \frac{\kappa}{\omega_{ph}^2 - \omega_1^2}}{1 - \frac{\kappa}{\omega_M^2 - \omega_2^2} \frac{\kappa}{\omega_{ph}^2 - \omega_1^2}} \tag{S46}$$

For the detected magnetic signals, the amplitudes of the two modes $B_{m1}$ and $B_{m2}$ are:

$$B_{m1} = B_1 = \frac{m_0 + \eta_0 \frac{\kappa}{\omega_M^2 - \omega_2^2}}{1 - \frac{\kappa}{\omega_M^2 - \omega_2^2} \frac{\kappa}{\omega_{ph}^2 - \omega_1^2}} \tag{S47}$$

$$B_{m2} = -\frac{\kappa}{\omega_M^2 - \omega_2^2} B_2 = -\frac{\kappa}{\omega_M^2 - \omega_2^2} \frac{\eta_0 + m_0 \frac{\kappa}{\omega_{ph}^2 - \omega_1^2}}{1 - \frac{\kappa}{\omega_M^2 - \omega_2^2} \frac{\kappa}{\omega_{ph}^2 - \omega_1^2}} \tag{S48}$$

The coupling coefficient $\kappa$:

$$\kappa = \gamma^2 \left( H_{ext} + 2H_{k0} \sin^2 \theta_H' \right) \frac{\sin(2\theta_H')}{2} \frac{3}{4} \frac{b_2}{M_s} = 272 \text{ GHz}^2 \tag{S49}$$

$M_s$ is the saturation magnetization. Note that $\kappa$ is obtained from the literature values

of fcc Co magnetostriction weighted by the Co percentage in the multilayer. The same theory and initial conditions are used to predict both amplitude and frequency.

Condition I:

When $\omega_M$ and $\omega_{ph}$ are far apart (as $\omega_M$ is tunable by the applied field, away from the coupling regime) and $\omega_M < \omega_{ph}$, then

$$4\kappa^2 \ll \left(\omega_M^2 - \omega_{ph}^2\right)^2 \quad \text{(S50)}$$

Thus, from Eq. (S31) and Eq. (S32)

$$\omega_1 = \omega_M \quad \text{(S51)}$$

$$\omega_2 = \omega_{ph} \quad \text{(S52)}$$

$$B_1 \approx m_0 + \eta_0 \frac{\kappa}{\omega_M^2 - \omega_{ph}^2} \quad \text{(S53)}$$

$$B_2 \approx \eta_0 + m_0 \frac{\kappa}{\omega_{ph}^2 - \omega_M^2} \quad \text{(S54)}$$

With this approximation, $m_{z'}(t)$ is simplified as:

$$m_{z'}(t) = -\frac{\kappa}{\omega_M^2 - \omega_2^2}\eta_0 \cos\omega_{ph}t + \left(m_0 + \eta_0 \frac{\kappa}{\omega_M^2 - \omega_2^2}\right)\cos\omega_M t \quad \text{(S55)}$$

The first term is the mixing mode arising from the magnon-phonon coupling. Its frequency is the phonon frequency $\omega_M > \omega_{ph}$ ($\approx 60$ GHz), independent of the magnetic field. Its amplitude is $\frac{\kappa}{\omega_M^2 - \omega_{ph}^2}\eta_0 \ll \eta_0$ $\frac{\kappa}{\omega_M^2 - \omega_{ph}^2}\eta_0 \ll \eta_0$, a small number because the coupling is very weak when $\omega_M$ and $\omega_{ph}$ are far apart.

The second term is the directly excited mode from the laser heating. Its frequency is the magnon frequency $\omega_M$, tunable by the magnetic field. Its amplitude is $\sim m_0$, a large amplitude and directly excited by the laser heating. Under such a condition, this term dominates the magnetic dynamics and is reflected as the "FMR" behavior in Fig. 4A.

Condition II:

When $\omega_M = \omega_{ph}$ (as obtained by tuning $\omega_M$ to a constant $\omega_{ph}$ via varying the magnetic field),

$$\omega_1 = \sqrt{\omega_{ph}^2 - \kappa} = \omega_{ph} - \frac{1}{2}\frac{\kappa}{\omega_{ph}} \quad (S56)$$

$$\omega_2 = \sqrt{\omega_{ph}^2 + \kappa} = \omega_{ph} + \frac{1}{2}\frac{\kappa}{\omega_{ph}} \quad (S57)$$

the difference between the two frequencies are:

$$\Delta\omega = \omega_2 - \omega_1 = \frac{\kappa}{\omega_{ph}} \quad (S58)$$

where $\Delta\omega$ is the frequency gap opened by the magnon-phonon coupling. With this approximation, $m_{z'}(t)$ is simplified as:

$$m_{z'}(t) = -\frac{\eta_0 - m_0}{2}\cos\omega_1 t + \frac{m_0 + \eta_0}{2}\cos\omega_2 t \quad (S59)$$

$$m_{z'}(t) = -\frac{\eta_0 - m_0}{2}\cos\sqrt{\omega_{ph}^2 - \kappa}\,t + \frac{m_0 + \eta_0}{2}\cos\sqrt{\omega_{ph}^2 + \kappa}\,t \quad (S60)$$

We define $\omega_M = \omega_{ph}$ as the resonance point. At the resonance point and nearby, the magnon-phonon coupling is strong. This produces two modes (Modes 1 and 2 in Fig. 3A), both are coupled magnon-phonon modes. The frequency gap between Modes 1 and 2 at the resonance regime depends on the strength of the coupling (a wider gap typically corresponds to a stronger magnon-phonon coupling). The hybridization regime is defined as the external field at which $|\omega_M - \omega_{ph}|$ approximates $\kappa/\omega_{ph}$, leading to a range of 19.7 kOe < $H_{ext}$ < 22.1 kOe. Their amplitudes are comparable and enhanced, compared to those under weak coupling conditions (see Fig. 3B).

## E. Strain-Assisted Switching Simulation

We find that magnetization switching can be realized in perpendicular materials with sufficiently large but reasonable strain amplitudes, shown in Figs. S9A and S9B. We found that if the 0.5% or 1.0% strain can be applied for this film, magnetization switching will occur resulting from the large-angle magnetization precession. The switching is fast, occurring within ~100 ps.

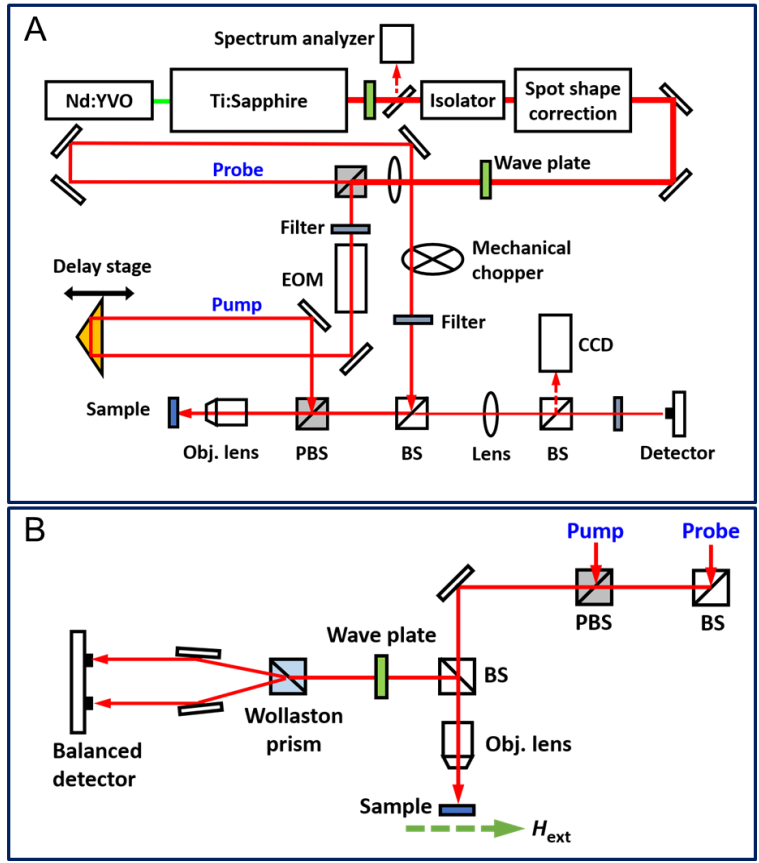

**Fig. S1. Schematics of setup platforms.** (**A**) The optical layout of the TDTR experimental setup. (**B**) The additional accessories for TR-MOKE measurements.

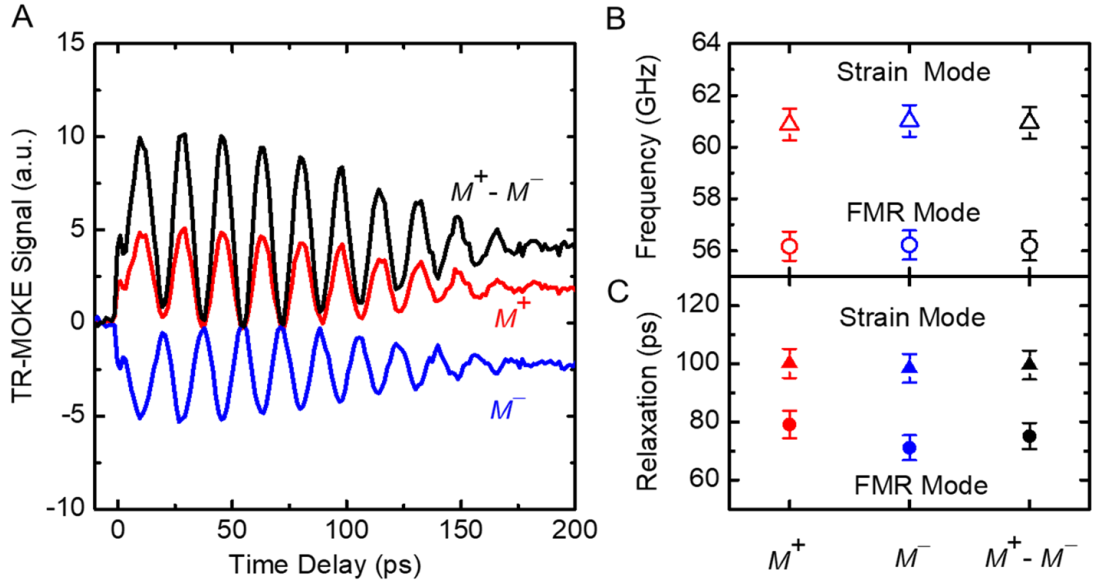

**Fig. S2. Verification of negligible non-MOKE contribution to TR-MOKE data. (A)** The TR-MOKE signals of a [Co(0.8 nm)/Pd(1.5 nm)]$_{11}$ multilayer near magneto-acoustic resonance. $M^+$ indicates the signal with a field of ~21 kOe applied at $\theta_H \sim 80°$ and $M^-$ indicates the signal with the magnetization in the opposite direction ($\theta_H \sim 260°$). $M^+ - M^-$ is the subtraction of both signals (to subtract non-MOKE components). **(B)**, **(C)** The extracted FMR and strain frequencies and relaxation times from the three different signals. Error bars are < 1% for the frequency (GHz) and < 5% for the relaxation time (ps), respectively.

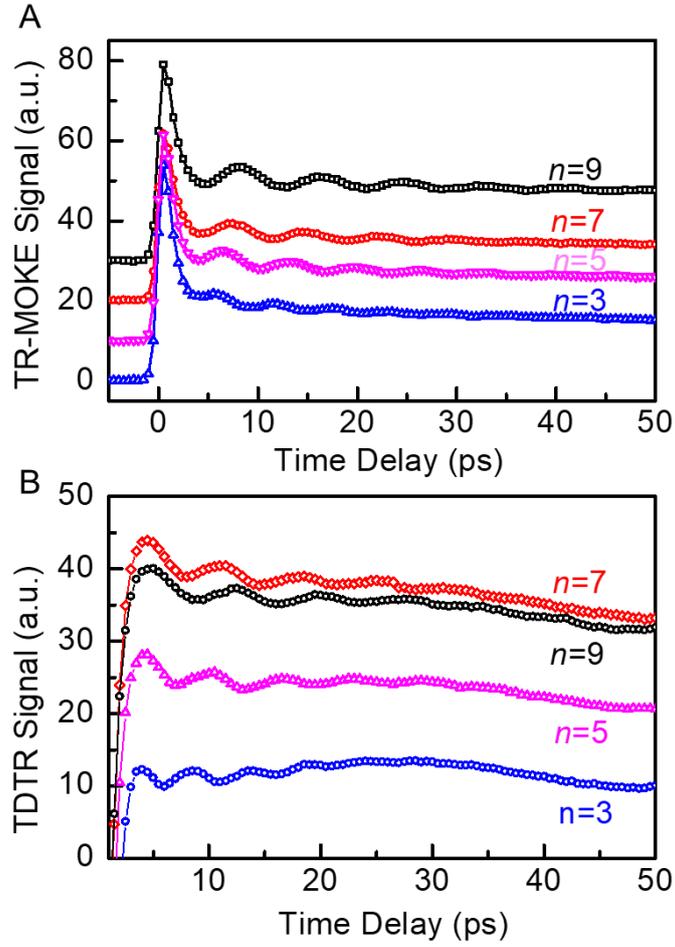

**Fig. S3. TR-MOKE data of the [Co(0.4 nm)/Pd(0.9 nm)]$_n$ multilayers with varying layer numbers.** (**A**), (**B**) TR-MOKE and TDTR signals of [Co(0.4 nm)/Pd(0.9 nm)]$_n$ ($n$ = 3, 5, 7, and 9) multilayers without $H_{ext}$, respectively. We can clearly see the oscillation of TR-MOKE and TDTR signals, implying that these samples possess the magnetostriction effect.

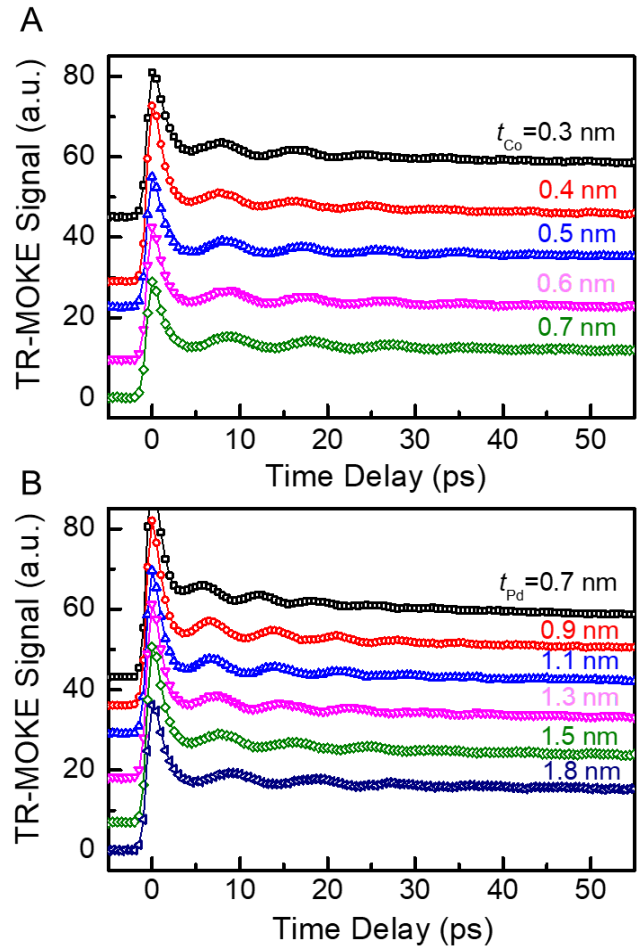

**Fig. S4. TR-MOEK data of the [Co(*x*)/Pd(*y*)]₇ multilayers with varying thicknesses of the Co and Pd layers.** (**A**), (**B**) TR-MOKE signals of [Co(*x*)/Pd(*y*)]₇ multilayers without $H_{ext}$ by changing the thickness of Co layer (*x* = 0.3 - 0.7 nm, *y* = 0.9 nm) and the thickness of Pd layer (*x* = 0.4 nm, *y* = 0.7 - 1.8 nm), respectively. From these results, we can deduce that the ASWs tilt the magnetization of the [Co/Pd]$_n$ multilayers and induce the magnetization oscillations.

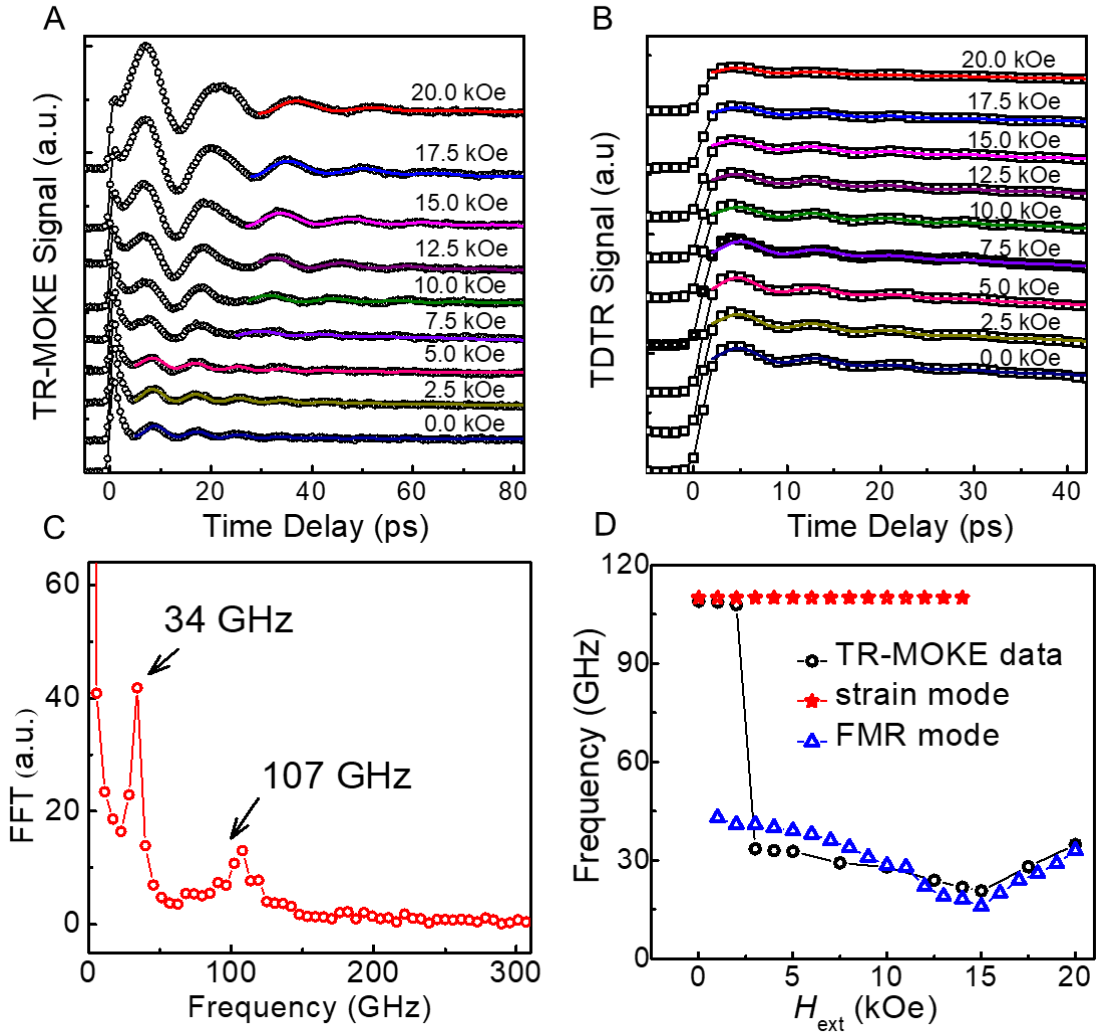

**Fig. S5. Existence of the strain and FMR modes in the [Co(0.7 nm)/Pd(1.5 nm)]$_7$ multilayer.** (**A**), (**B**) The experimental and fitting TR-MOKE and TDTR signals of the [Co(0.7 nm)/Pd(1.5 nm)]$_7$ multilayer under in-plane $H_{ext}$ ranging from 0-20 kOe. (**C**) Fourier transform on the TR-MOKE signal under 3 kOe in-plane $H_{ext}$, from which both the FMR and strain mode peaks (34 GHz of spin damping and 107 GHz of acoustic strain wave) can be found. (**D**) The separated frequencies (FMR and strain modes) fitted from the TR-MOKE signals. We found the frequency of strain-like mode keeps the constant ~107 GHz, however, the frequency of FMR mode precession first decreases then increases up to 30 GHz, which could be up to 107 GHz with larger $H_{ext}$.

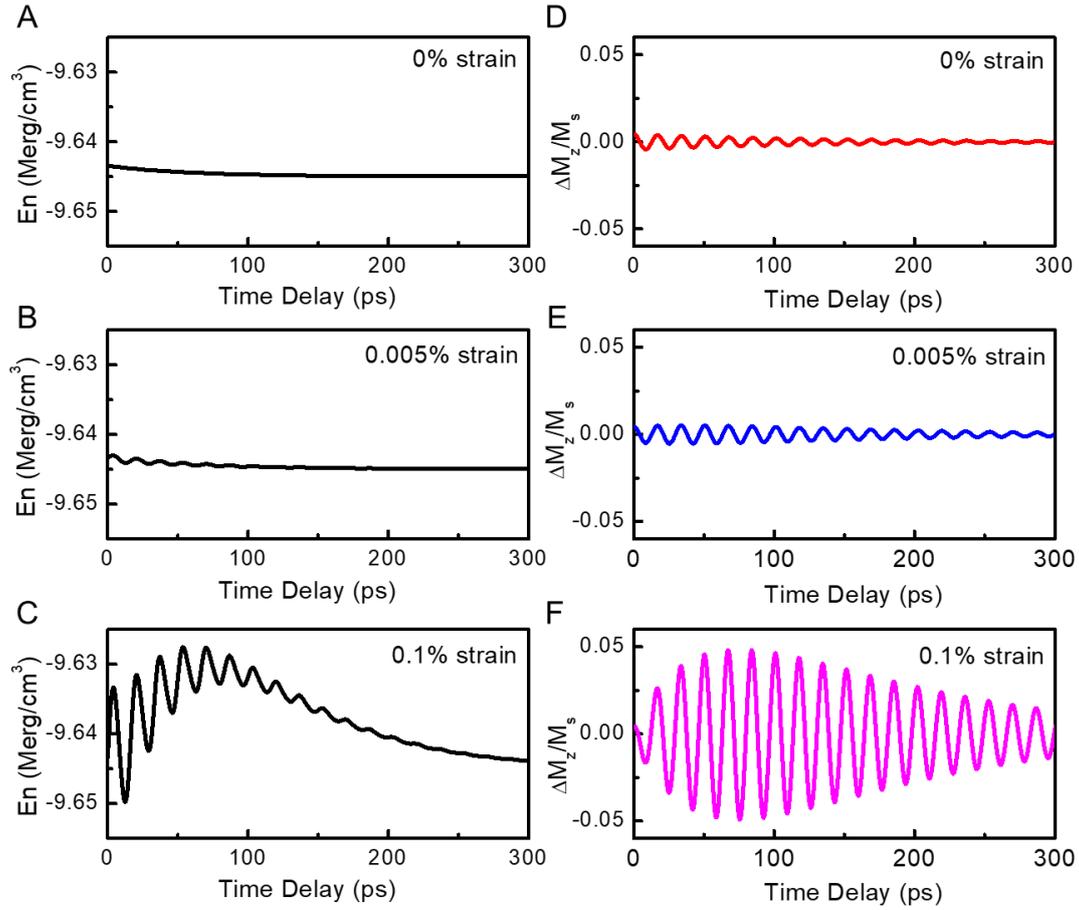

**Fig. S6. Relationship between energy density and resonance.** (**A-C**) The time evolution of the energy density of the spin system for 0%, 0.005% and 0.1% strains, respectively. (**D-F**) The time evolution of the out-of-plane magnetization component ($\Delta M_z/M_s$) for 0%, 0.005% and 0.1% strains with the $H_{ext}$ of 21 kOe, respectively. It is very clear that when the 0% strain is applied, $\Delta M_z/M_s$ shows the damped-like oscillations. At the moderate strain such as 0.005%, the spins show a constant amplitude at the beginning 100 ps. At the relatively large 0.1% strain, $\Delta M_z/M_s$ presents the magneto-acoustic resonance behavior. This is due to the relationship between the energy pumped from strain, and the energy dissipated from the system determined by the damping factor.

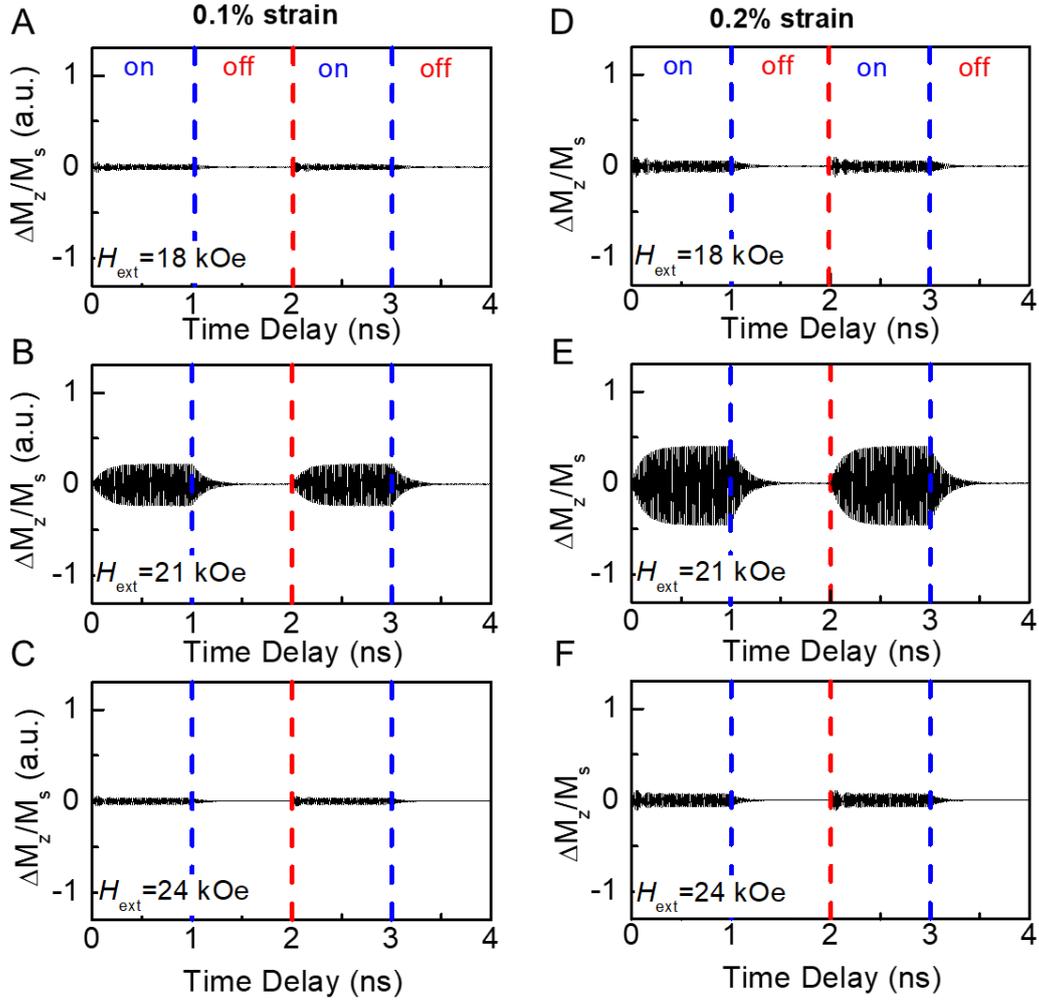

**Fig. S7. Manipulation of resonance through external magnetic field.** The spin precession coupled with (**A-C**) 0.1% and (**D-F**) 0.2% strain pulses under $H_{ext}$ of 18 kOe, 21 kOe and 24 kOe, respectively. The material is [Co(0.7 nm)/Pd (1.5 nm)]$_7$ film using the measured magnetic properties. When the strain pulse is on, the system gets excited rapidly to an enhanced large angle precession with a rise time of ~ 200 ps under $H_{ext}$ = 21 kOe. When the strain pulse is off, the system at all three $H_{ext}$ values shows relaxation behavior.

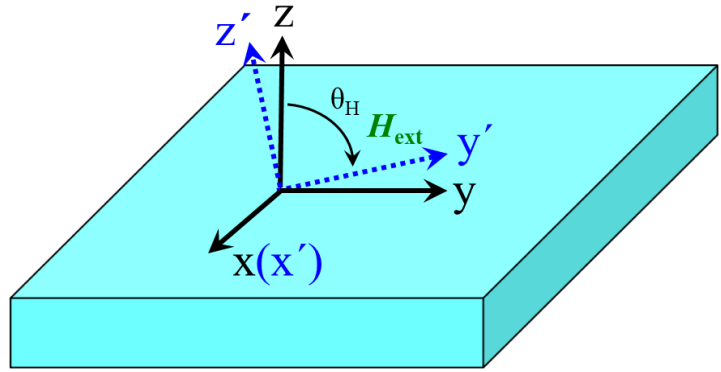

**Fig. S8. Coordinate transformation.** $x, y, z$ are the thin film coordinate in which the $z$ axis is the normal vector of the thin film. $x', y', z'$ are the field coordinate in which the $y'$ axis is the external magnetic field ($H_{ext}$) direction.

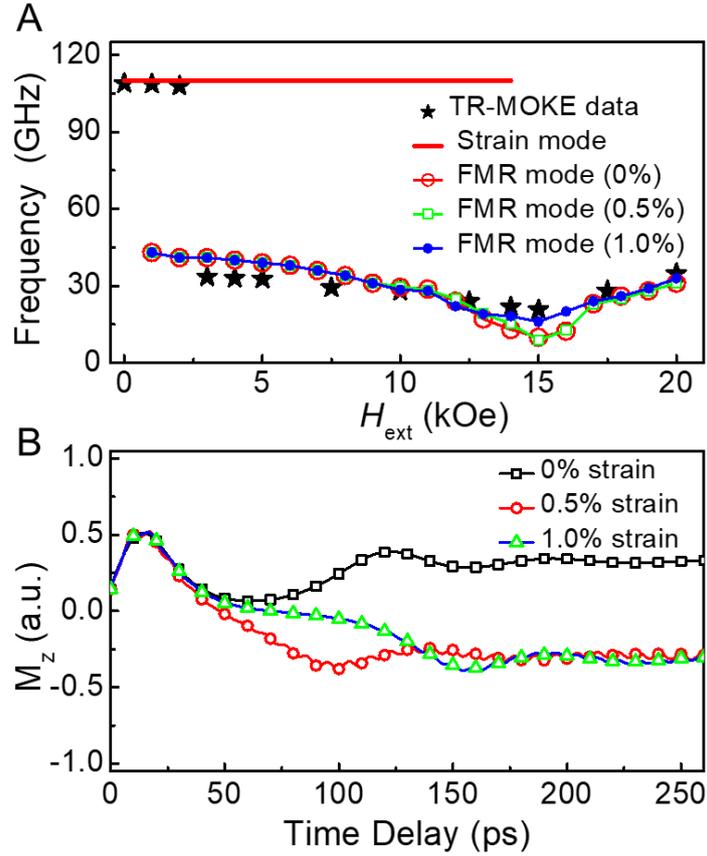

**Fig. S9. Strain-driven magnetization switching.** (**A**) The experimental TR-MOKE (black circle) and simulated frequencies [strain-like mode: red circle; FMR mode (0% strain): blue triangle; FMR mode (0.5% strain): blue triangle; FMR mode (1.0% strain): green diamond] as a function of $H_{ext}$ for the [Co(0.7 nm)/Pd(1.5 nm)]$_7$ multilayer. The strain mode corresponds to the frequency of ASWs from TDTR fitting and the FMR mode denotes the frequency of spin precession with the different strains. (**B**) The change of $M_z$ with the different strains [0% strain (black square), 0.5% strain (red circle), and 1% strain (blue triangle)] as a function of the time delay for the [Co(0.7 nm)/Pd(1.5 nm)]$_7$ multilayer with $H_{ext}$ = 15 kOe. It is found that the magnetization could be switched when the 0.5% and 1.0% strains are applied.

48